# The Search for Stable Nickel: Investigating the Origins of Type Ia Supernovae with Late-time Near-infrared Spectroscopy from the Carnegie Supernova Project-II

Sahana Kumar[1], Eric Y. Hsiao[2], Chris Ashall[3], Peter Hoeflich[2], Eddie Baron[4,5], Mark M. Phillips[6], Maryam Modjaz[1], Abigail Polin[7], Nidia Morrell[6], Christopher R. Burns[8], Jing Lu[9], Melissa Shahbandeh[10], Lindsey A. Kwok[11], Lluis Galbany[12,13], M. D. Stritzinger[14], Carlos Contreras[6], James M. DerKacy[10], T. Hoover[7], Syed A. Uddin[15,16], Saurabh W. Jha[17], Huangfei Xiao[2], K. Krisciunas[18], and N. B. Suntzeff[18]

[1] Department of Astronomy, University of Virginia, 530 McCormick Rd., Charlottesville, VA 22904, USA; sahanak@gmail.com
[2] Department of Physics, Florida State University, 77 Chieftan Way, Tallahassee, FL 32306, USA
[3] Institute for Astronomy, University of Hawai'i, Honolulu, HI 96822, USA
[4] Planetary Science Institute, 1700 East Fort Lowell Rd., Suite 106, Tucson, AZ 85719-2395, USA
[5] Hamburger Sternwarte, Gojenbergsweg 112, D-21029 Hamburg, Germany
[6] Carnegie Observatories, Las Campanas Observatory, Colina El Pino, Casilla 601, Chile
[7] Department of Physics and Astronomy, Purdue University, 525 Northwestern Ave., West Lafayette, IN 47907, USA
[8] Observatories of the Carnegie Institution for Science, 813 Santa Barbara St., Pasadena, CA 91101, USA
[9] Department of Physics and Astronomy, Michigan State University, East Lansing, MI 48824, USA
[10] Space Telescope Science Institute, 3700 San Martin Drive, Baltimore, MD 21218, USA
[11] Center for Interdisciplinary Exploration and Research in Astrophysics (CIERA), 1800 Sherman Ave., Evanston, IL 60201, USA
[12] Institute of Space Sciences (ICE, CSIC), Campus UAB, Carrer de Can Magrans, s/n, E-08193 Barcelona, Spain
[13] Institut d'Estudis Espacials de Catalunya (IEEC), E-08034 Barcelona, Spain
[14] Department of Physics and Astronomy, Aarhus University, Ny Munkegade 120, DK-8000 Aarhus C, Denmark
[15] American Public University System, 111 W. Congress St., Charles Town, WV 25414, USA
[16] Center for Astronomy, Space Science and Astrophysics, Independent University, Bangladesh, Bashundhara R/A, Dhaka 1245, Bangladesh
[17] Department of Physics and Astronomy, Rutgers, the State University of New Jersey, 136 Frelinghuysen Road, Piscataway, NJ 08854-8019, USA
[18] George P. and Cynthia Woods Mitchell Institute for Fundamental Physics and Astronomy, Department of Physics and Astronomy, Texas A&M University, College Station, TX 77843, USA

## Abstract

Producing stable $^{58}$Ni in Type Ia supernovae (SNe Ia) requires sufficiently high-density conditions that are not predicted for all origin scenarios, so examining the distribution of $^{58}$Ni using the near-infrared (NIR) [Ni II] 1.939 $\mu$m line may observationally distinguish between possible progenitors and explosion mechanisms. We present 79 telluric-corrected NIR spectra of 22 low-redshift SNe Ia from the Carnegie Supernova Project-II, ranging from +50 to +505 days, including 31 previously unpublished spectra. We introduce the Gaussian Peak Ratio, a detection parameter that confirms the presence of the NIR [Ni II] 1.939 $\mu$m line in eight SNe in our sample. Nondetections occur at earlier phases ($\leqslant$+100 days) when the NIR Ni line has not emerged yet or in low signal-to-noise spectra, yielding inconclusive results. Subluminous 86G-like SNe Ia show the earliest NIR Ni features around ∼+50 days, whereas normal-bright SNe Ia do not exhibit NIR Ni until ∼+150 days. NIR Ni features detected in our sample have low peak velocities ($v \sim 1200$ km s$^{-1}$) and narrow line widths ($\leqslant$3500 km s$^{-1}$), indicating stable $^{58}$Ni is centrally located. This implies high-density burning conditions in the innermost regions of SNe Ia and could be due to higher mass progenitors (i.e., near-$M_{ch}$). NIR spectra of the nearly two dozen SNe Ia in our sample are compared to various model predictions and paired with early-time properties to identify ideal observation windows for future SNe Ia discovered by upcoming surveys with Rubin-LSST or the Roman Space Telescope.

## 1. Introduction

Type Ia supernovae (SNe Ia) are a well-known tool for cosmology and have been used to study the expansion history of our universe (A. G. Riess et al. 1998; S. Perlmutter et al. 1999). Empirical relations allow SNe Ia to be used as standardizable candles (M. M. Phillips 1993), but it is currently uncertain if these correlations evolve with redshift, and understanding the details of SNe Ia is crucial for improving precision cosmology. SNe Ia are homogeneous compared to other astrophysical phenomena, but there is still variation in the photometric and spectroscopic properties of the normal SN Ia population (e.g., S. W. Jha et al. 2019; A. J. Ruiter & I. R. Seitenzahl 2025). This diversity introduces uncertainties that may hinder the improvement of cosmological measurements (M. Betoule et al. 2014).

Based on observed luminosities and nucleosynthetic yields, SNe Ia are understood to be the thermonuclear explosion of at least one carbon–oxygen (C/O) white dwarf (WD; F. Hoyle & W. A. Fowler 1960), but the observed differences between various subtypes (e.g., S. Taubenberger 2017; S. W. Jha et al. 2019) question whether all SNe Ia are produced by the same physical process. In theory, the progenitor system can either be a single-degenerate system with one WD and a nondegenerate companion star, or a double-degenerate system composed of two WDs. Prevalent explosion mechanisms for SNe Ia include delayed detonation (DDT; A. Khokhlov et al. 1992), helium detonation (also known as double detonation; S. E. Woosley & T. A. Weaver 1994), and variations on mergers or collisions of WDs (S. Rosswog et al. 2009).

Although progenitor systems and explosion mechanisms are two distinct topics, understanding the physical properties of the progenitor WD can provide further information on possible explosion mechanisms. The mass of the progenitor WD is currently a very active topic of debate and may hold the key to understanding intrinsic diversity within the SN Ia population. Late-time spectroscopy can be used to detect signatures of specific explosion mechanisms and provide clues on the state of the progenitor WD shortly before its demise. With the current debate on the origins of SNe Ia, independent measurements of progenitor properties can be crucial discriminators.

The central density is a strong indicator of the mass of the exploding WD, and observables associated with progenitor central density are potentially capable of distinguishing between the explosions of a sub-Chandrasekhar-mass (sub-$M_{ch}$) or near-Chandraskehar-mass ($M_{ch}$) WD. Lower densities in the progenitor's innermost regions are a natural consequence of the lower total masses of sub-$M_{ch}$ WDs. Decreasing the density can decrease the efficiency of electron capture in the central region (P. Höflich et al. 2004) and often results in sub-$M_{ch}$ models producing less stable material than $M_{ch}$ models (e.g., S. Blondin et al. 2018; K. D. Wilk et al. 2018; L. J. Shingles et al. 2020). To produce significant amounts of stable iron group elements (IGEs), sufficiently high densities and temperatures are required, which are not predicted for all explosion mechanisms (F. Lach et al. 2020; R. Pakmor et al. 2024).

Stable nickel in the form of $^{58}$Ni is uniquely useful for studying the production of stable material, thanks to the relatively short half-life of radioactive isotope $^{56}$Ni. By the time the SN ejecta becomes optically thin enough to observe the innermost regions, the majority of radioactive $^{56}$Ni has decayed away, and thus any Ni emission is due to stable isotopes such as $^{58}$Ni. Radioactive isotopes of other IGEs, such as $^{56}$Co and $^{57}$Fe, have much longer half-lives (ranging from several months to several years), and some are daughter products of $^{56}$Ni decay, so it is more difficult to observationally isolate the stable material synthesized in the explosion. Stable $^{58}$Ni is also a product of high-density nuclear burning (F. K. Thielemann et al. 1986), so the location and distribution of stable $^{58}$Ni within the SN ejecta can be used to test predictions from explosion models.

Given the strong promise of stable Ni's constraining power for SN Ia explosion models, there have been many attempts to observationally detect its presence in the optical, near-infrared (NIR), and mid-infrared (MIR). The optical [Ni II] $\lambda$7378 Å emission line is a characteristic feature of nebular-phase SNe Ia, but significant line blending with neighboring [Fe II] $\lambda\lambda$7155, 7172, 7388, 7453 and possibly [Ca II] $\lambda\lambda$7291, 7324 lines can limit accurate determination of emission line parameters. Stable Ni has also been detected in the MIR using the Spitzer Space Telescope (SST; e.g., C. L. Gerardy et al. 2007; C. M. Telesco et al. 2015) and the James Webb Space Telescope (e.g., J. M. DerKacy et al. 2023; L. A. Kwok et al. 2023; C. Ashall et al. 2024). Very few objects have published MIR spectra obtained with space telescopes, and depending on the resolution of the observed spectrum, some narrow emission features can be unresolved (J. M. DerKacy et al. 2024). Furthermore, uncertainties in atomic data can overestimate line strengths for higher ionization states, thus limiting the possible use of some MIR Ni features to examine physical conditions within the SN ejecta (S. Blondin et al. 2023).

The NIR currently provides one of the best opportunities to study the presence of stable Ni for a larger sample of SNe Ia. The longer wavelength means the NIR goes "nebular" (i.e., optically thin) before the optical (J. C. Wheeler et al. 1998), potentially providing earlier opportunities to observe the SN's inner regions. When considering different subtypes of SNe Ia at similar phases, NIR spectra often show more drastic variation compared to optical spectra (see Figure 6 of E. Y. Hsiao et al. 2019).

The seminal work of B. Friesen et al. (2014) identified the NIR [Ni II] $\lambda$1.939 $\mu$m line by comparing observed NIR spectra of SNe Ia to synthetic spectra generated using various combinations of permitted and forbidden lines. B. Friesen et al. (2014) found that this NIR [Ni II] emission line may be observable as early as $\sim$+50 days post-peak brightness. In general, spectral models predict the [Ni II] $\lambda$1.939 $\mu$m line to be the strongest Ni line in the NIR (A. Flörs et al. 2020; P. Hoeflich et al. 2021; S. Blondin et al. 2023) and better-isolated than its optical counterparts (B. Friesen et al. 2014; K. D. Wilk et al. 2018). Furthermore, this emission line is notably absent in synthetic spectra from multiple sub-$M_{ch}$ models (e.g., S. Blondin et al. 2018), and thus may provide a way of discerning between $M_{ch}$ and sub-$M_{ch}$ explosions.

Although the NIR [Ni II] $\lambda$1.939 $\mu$m line has been predicted and modeled for many years (e.g., B. Friesen et al. 2014), observing that wavelength region using ground-based telescopes is challenging due to strong telluric absorption caused by water vapor in the atmosphere. Previous observational studies of the 1.9 $\mu$m region in SNe Ia at nebular phases have been limited to a handful of bright, nearby objects (e.g., B. Friesen et al. 2014; S. Blondin et al. 2015; S. Dhawan et al. 2018).

Here, we present the largest sample of high signal-to-noise (S/N) late-time NIR spectra of SNe Ia published to date, with 79 spectra of 22 SNe Ia, including 31 previously unpublished nebular-phase NIR spectra obtained as a part of the Carnegie Supernova Project-II (CSP-II; 2019; M. M. Phillips et al. 2019; E. Y. Hsiao et al. 2019). When combined with previously published NIR spectra, this sample extends time-series spectroscopy for 13 nearby SNe Ia. This work examines the 1.9 $\mu$m region for the possible presence of the [Ni II] 1.939 $\mu$m feature using a uniform sample of high-S/N and telluric-corrected NIR spectra at late times.

With spectra ranging from +50 to +505 days post-peak brightness, this dataset provides an excellent opportunity to study the late-time NIR behavior of nearly two dozen low redshift SNe Ia. For the purposes of this work, the "transitional phase" refers to $\sim$+50 to $\sim$+100 days, and any epochs past +100 days are considered to be during the "nebular phase." The substantial phase range of the NIR spectra presented in this work allows us to study the time evolution of specific observables for a low redshift sample of SNe Ia. This information can be used to pair complementary early- and late-time properties to make more efficient use of observing facilities in the future.

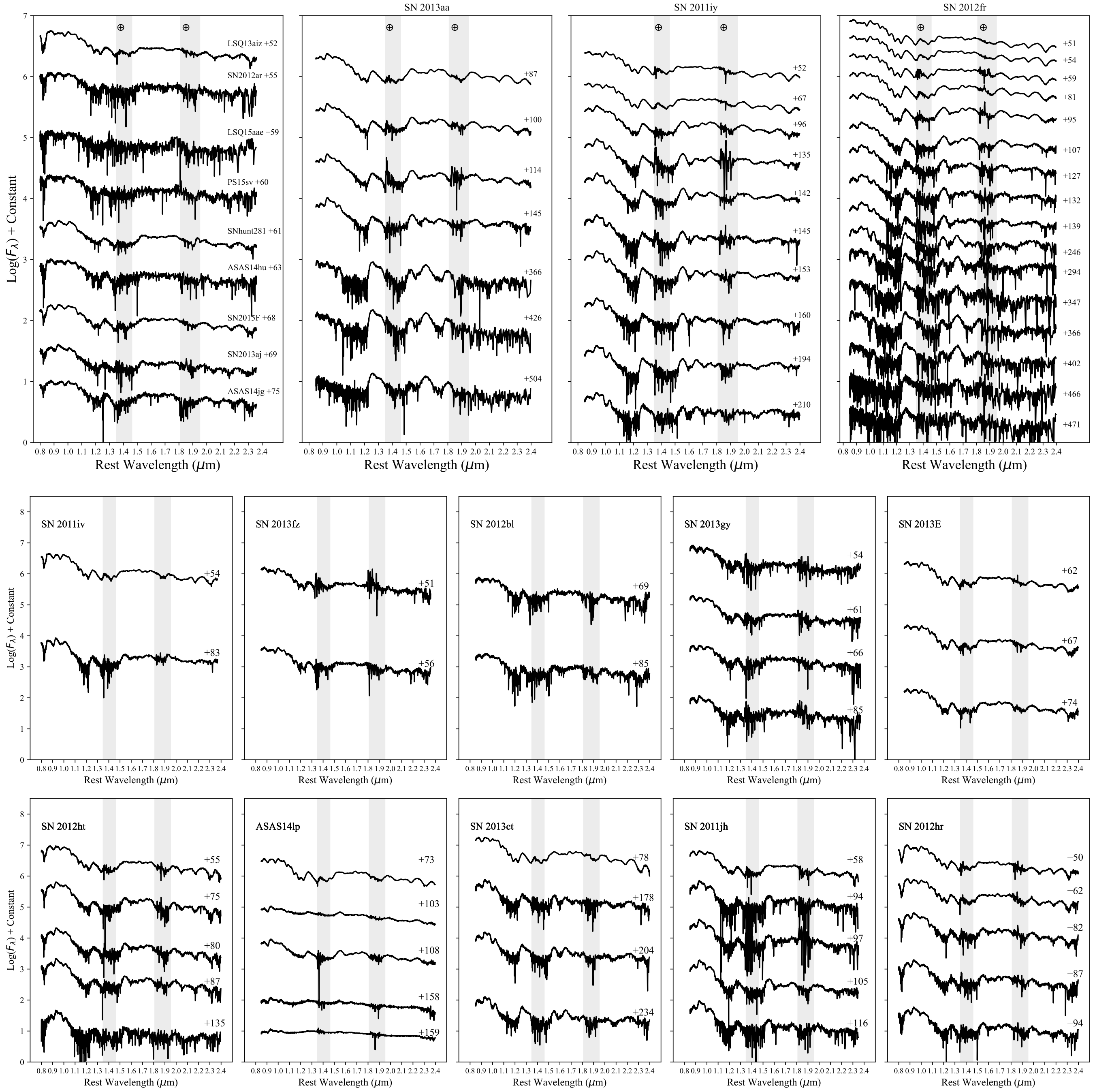

**Figure 1.** All NIR spectra used in this study. Telluric bands are the gray regions, and the top left panel shows the objects in this dataset that have only one late-time spectrum observed with FIRE by CSP-II. SN 2012fr has the largest time-series spectroscopic dataset, and SN 2013aa has the latest spectrum at +505 days post-peak brightness. Further details on the spectroscopic observations are presented in Table 1 and Section 2. The 80 spectra are available in FITS format in a .tar.gz package as the data behind the figure.

(The data used to create this figure are available in the online article.)

## 2. Observations

One of the strengths of this dataset is the homogeneity of the observations. All NIR spectra in this sample were obtained using the same instrument, observing strategy, and data reduction pipeline. A telluric standard was observed for each NIR spectrum, and telluric corrections were applied using a consistent method.

The following sections provide further details on the NIR data used in this work. Section 2.1 summarizes the spectroscopic observations, followed by a description of telluric corrections in Section 2.2. Finally, Section 2.3 highlights relevant photometric parameters of the SNe Ia in this sample. All NIR spectra used in this study are presented in Figure 1, including 31 previously unpublished nebular-phase spectra.

### 2.1. Magellan–Baade + FIRE

All NIR spectra in this sample were observed using the FIRE (R. A. Simcoe et al. 2013) spectrograph mounted on the 6.5 m Magellan–Baade Telescope at Las Campanas Observatory in Chile. These spectra were observed as part of the Carnegie Supernova Project-II (CSP-II; E. Y. Hsiao et al. 2019; M. M. Phillips et al. 2019), a four-year observing program operating from 2011 to 2015 that obtained optical and NIR observations of 214 SNe Ia, including a low redshift subsample ($z \leqslant 0.04$) collected to improve our understanding of SNe Ia physics. The CSP-II NIR spectroscopy program has published first-of-its-kind datasets for early-time SNe Ia (J. Lu et al. 2023), SNe II (S. Davis et al. 2019), and SESNe (M. Shahbandeh et al. 2022). All light-curve parameters used in this work were obtained using SNooPy (C. R. Burns et al. 2011) and are published in S. A. Uddin et al. (2024).

The FIRE spectra presented in this work were observed using the high-throughput prism mode paired with the $0\overset{''}{.}6$ slit, which results in a wavelength coverage of 0.8–2.5 $\mu$m. This results in resolutions of $R \sim 500$ in the $J$ band, $R \sim 450$ in the $H$ band, and $R \sim 300$ in the $K$ band. All spectroscopic observations were obtained using the conventional ABBA "nod-along-the-slit" method. The "sampling-up-the-ramp" readout mode was used for the science exposures to minimize the readout noise by sampling multiple nondestructive reads. Ne and Ar arc lamps were used for wavelength calibration, and the NIR spectra were reduced using the IDL pipeline firehose (R. A. Simcoe et al. 2013). Further details and an overview of CSP-II NIR spectroscopy for all types of SNe can be found in E. Y. Hsiao et al. (2019).

### 2.2. Telluric Corrections

Accurate telluric corrections are essential for this analysis because the [Ni II] 1.939 $\mu$m is in a wavelength region often obscured in ground-based observations by telluric absorption from the atmosphere. If a SN Ia has a very high redshift, it is possible that the feature may be at an observed wavelength red enough to avoid the telluric band from $\sim$1.81 to $\sim$1.95 $\mu$m. However, SNe Ia become faint very quickly, and high redshift objects become unobservable even sooner than their low redshift counterparts. This type of study can currently only be achieved with low redshift SNe Ia that can be observed at least during the transitional phase and preferably until the nebular phases.

We individually telluric corrected each NIR spectrum using the IDL tool xtellcor (W. D. Vacca et al. 2003). The same method is employed for all FIRE spectra in the CSP-II NIR spectroscopy dataset (E. Y. Hsiao et al. 2019). An A0V telluric standard star was observed with FIRE using the same instrument configuration as the SN observations and at a similar airmass. The A0V star spectrum is then matched to a model spectrum of Vega to identify and remove Paschen and Brackett Hydrogen lines from the observed spectrum. The A0V telluric standard star is also used for flux calibration.

Although A-type stars are particularly useful as telluric standards since their spectra contain fewer features than other spectral types and are well-approximated by a blackbody with a $T \sim 10{,}000$ K, we opt to use the relative flux of the NIR Ni II feature in most of our analysis as opposed to the absolute flux because the telluric correction method relies on optical $B$- and $V$-band magnitudes to extrapolate the NIR magnitudes (W. D. Vacca et al. 2003; E. Y. Hsiao et al. 2019). Table 1 lists the telluric standard star used for each NIR spectrum included in this work.

S. Davis et al. (2019) tested the quality of the CSP-II NIR telluric corrections by examining an SN II Paschen-$\alpha$ feature located within the telluric region between the $H$ and $K$ bands. This feature was detected with at least 10% precision in 70% of spectra in the CSP-II Type II SN sample, showcasing the high instrument sensitivity and throughput that allows for recovery of the SN spectrum in regions obscured by telluric features. This demonstrates the consistency and accuracy of telluric corrections applied to CSP-II FIRE spectra.

The combination of high-S/N observations and careful telluric corrections allows for analysis of the 1.9 $\mu$m region in a larger sample of SNe Ia. Of the 79 telluric-corrected spectra in our sample, over 70% have a $S/N \geqslant 5$ in the region between 1.8 and 2.1 $\mu$m. This results in 58 late-time NIR spectra of 18 low redshift SNe Ia with sufficiently high S/N in the telluric region to measure emission line parameters such as line velocity and line width. The measurement method applied to these spectra is described in Section 3.1, and the results are reported in Section 4.

### 2.3. CSP-II Light-curve Parameters

Nebular phase analyses often require information from earlier observations, such as light-curve parameters and subtype classification. The widespread use of automated transient surveys over the past decade has virtually guaranteed ground-based optical light curves for many new SNe. In order to plan follow-up spectroscopy based on discoveries from automated transient searches, we must identify useful photometric properties. Here, we present relevant optical light-curve parameters that will be used in the analysis for the SNe Ia in this sample.

Since it measures how quickly the light curve evolves, the light-curve decline-rate parameter ($\Delta m_{15}(B)$; M. M. Phillips 1993) is commonly used to discern between different subtypes of thermonuclear supernovae (e.g., S. Taubenberger 2017). However, there is a degeneracy in $\Delta m_{15}(B)$ values for SNe Ia with faster declining light curves if the light-curve shape changes to a linear decline before 15 days post-peak brightness (M. M. Phillips 2012; C. R. Burns et al. 2014). For this reason, many studies that consider SNe Ia at the rapidly declining end of the Phillips relation (i.e., K. Krisciunas et al. 2017; C. Ashall et al. 2019; O. Graur 2024) use the color-stretch parameter.

The color-stretch parameter, $s_{BV}$, is based on the timing between the $B$-band maximum and the peak of the $B-V$ color curve when the SN Ia is at its reddest color (C. R. Burns et al. 2014). SNe Ia with faster declining light curves are typically less luminous and have lower $s_{BV}$ values, whereas slowly declining SNe Ia have higher $s_{BV}$ values and are often the more luminous events within the SN Ia population. Although $s_{BV}$ is the specific light-curve parameter used in this study, color-stretch parameters can also be defined for other pairs of observing filters (C. Ashall et al. 2020), which is complementary to the multiband rolling cadence observing strategy adopted by future large surveys (e.g., F. B. Bianco et al. 2019).

## 3. [Ni II] $\lambda$1.939 $\mu$m Line Measurement Method

Although this nebular phase NIR Ni feature is predicted to be less blended than its optical counterparts at late times (K. D. Wilk et al. 2018), the large time coverage of this sample means we cannot assume zero line blending at earlier (i.e., transitional) phases.

Transitional phase spectra contain a combination of allowed and forbidden emission lines (G. H. Marion et al. 2009; E. E. Gall et al. 2012), which can make modeling this phase

**Table 1**
Spectroscopic Observations with the FIRE Spectrograph on the Magellan–Baade Telescope for the 79 NIR Spectra presentedin This Work

| SN | Phase (days) | UT Date (MM/DD/YY) | MJD | Exp Time (s) | Telluric Standard | References |
|---|---|---|---|---|---|---|
| ASAS14hu | 63 | 12/8/14 | 56999.08 | 1268.0 | HD 54682 | J. Lu et al. (2023) |
| ASAS14jg | 75 | 1/14/15 | 57036.04 | 1902.0 | HD 1881 | J. Lu et al. (2023) |
| ASAS14lp | 73 | 3/7/15 | 57088.31 | 285.3 | HD 97919 | J. Lu et al. (2023) |
| ASAS14lp | 103 | 4/7/15 | 57119.26 | 507.2 | HD 108140 | This work |
| ASAS14lp | 108 | 4/12/15 | 57124.20 | 507.2 | HD 97919 | This work |
| ASAS14lp | 158 | 6/1/15 | 57174.17 | 1014.4 | HD 105992 | This work |
| ASAS14lp | 159 | 6/2/15 | 57175.17 | 1268.0 | HD 105992 | This work |
| 2011iv | 54 | 2/3/12 | 55960.11 | 1200.0 | HD 97282 | C. Gall et al. (2018) |
| 2011iv | 83 | 3/3/12 | 55989.08 | 1014.4 | HD 27166 | C. Gall et al. (2018) |
| 2011iy | 52 | 1/19/12 | 55945.39 | 285.3 | HD 118997 | J. Lu et al. (2023) |
| 2011iy | 67 | 2/3/12 | 55960.26 | 1200.0 | HD 125062 | J. Lu et al. (2023) |
| 2011iy | 96 | 3/3/12 | 55989.29 | 2536.0 | HD 110649 | J. Lu et al. (2023) |
| 2011iy | 135 | 4/11/12 | 56028.25 | 1902.0 | HD 157334 | This work |
| 2011iy | 142 | 4/18/12 | 56035.27 | 1050.0 | HD 115414 | This work |
| 2011iy | 145 | 4/21/12 | 56038.29 | 900.0 | HD 112836 | This work |
| 2011iy | 154 | 4/30/12 | 56047.23 | 3170.0 | HD 117821 | This work |
| 2011iy | 160 | 5/7/12 | 56054.17 | 3804.0 | HD 118997 | This work |
| 2011iy | 194 | 6/10/12 | 56088.12 | 2536.0 | HD 117248 | This work |
| 2011iy | 210 | 6/26/12 | 56104.15 | 2536.0 | HD 117248 | This work |
| 2011jh | 58 | 3/3/12 | 55989.24 | 1268.0 | HD 121884 | J. Lu et al. (2023) |
| 2011jh | 94 | 4/8/12 | 56025.10 | 2874.4 | HD 123426 | J. Lu et al. (2023) |
| 2011jh | 97 | 4/11/12 | 56028.13 | 2536.0 | HD 105992 | J. Lu et al. (2023) |
| 2011jh | 105 | 4/19/12 | 56036.20 | 3600.0 | HD 110749 | This work |
| 2011jh | 116 | 4/30/12 | 56047.16 | 3170.0 | HD 97919 | This work |
| 2012ar | 55 | 4/30/12 | 56047.31 | 3804.0 | HD 157334 | J. Lu et al. (2023) |
| 2012bl | 69 | 6/10/12 | 56088.26 | 1743.5 | HD 171792 | J. Lu et al. (2023) |
| 2012bl | 85 | 6/26/12 | 56104.40 | 2377.5 | HD 203999 | J. Lu et al. (2023) |
| 2012fr | 52 | 1/3/13 | 56295.18 | 507.2 | HD 24083 | J. Lu et al. (2023) |
| 2012fr | 55 | 1/6/13 | 56298.10 | 1014.4 | HD 24475 | J. Lu et al. (2023) |
| 2012fr | 60 | 1/11/13 | 56303.03 | 507.2 | HD 24083 | J. Lu et al. (2023) |
| 2012fr | 82 | 2/2/13 | 56325.03 | 507.2 | HD 19839 | J. Lu et al. (2023) |
| 2012fr | 96 | 2/16/13 | 56339.02 | 634.0 | HD 33715 | J. Lu et al. (2023) |
| 2012fr | 108 | 2/28/13 | 56351.07 | 634.0 | HD 24475 | This work |
| 2012fr | 128 | 3/20/13 | 56371.00 | 1268.0 | HD 23722 | This work |
| 2012fr | 133 | 3/25/13 | 56376.01 | 1014.4 | HD 23722 | This work |
| 2012fr | 140 | 4/1/13 | 56383.01 | 1268.0 | HD 24475 | This work |
| 2012fr | 246 | 7/16/13 | 56489.38 | 2694.5 | HD 26760 | This work |
| 2012fr | 294 | 9/2/13 | 56537.38 | 507.2 | HD 14003 | This work |
| 2012fr | 347 | 10/25/13 | 56590.35 | 3804.0 | HIP 23691 | This work |
| 2012fr | 367 | 11/14/13 | 56610.27 | 1585.0 | HD 26760 | This work |
| 2012fr | 403 | 12/20/13 | 56646.24 | 1521.6 | HD 19839 | This work |
| 2012fr | 467 | 2/22/14 | 56710.11 | 1268.0 | HD 41171 | This work |
| 2012fr | 472 | 2/27/14 | 56715.09 | 951.0 | HD 243374 | This work |
| 2012hr | 50 | 2/16/13 | 56339.03 | 507.2 | HD 33715 | J. Lu et al. (2023) |
| 2012hr | 62 | 2/28/13 | 56351.10 | 507.2 | HD 47518 | J. Lu et al. (2023) |
| 2012hr | 82 | 3/20/13 | 56371.04 | 1014.4 | HD 47518 | J. Lu et al. (2023) |
| 2012hr | 87 | 3/25/13 | 56376.04 | 1014.4 | HD 47518 | J. Lu et al. (2023) |
| 2012hr | 94 | 4/1/13 | 56383.07 | 1014.4 | HD 47518 | J. Lu et al. (2023) |
| 2012ht | 55 | 2/28/13 | 56351.23 | 760.8 | HIP 64248 | J. Lu et al. (2023) |
| 2012ht | 75 | 3/20/13 | 56371.11 | 634.0 | HD 96781 | J. Lu et al. (2023) |
| 2012ht | 80 | 3/25/13 | 56376.12 | 760.8 | HD 96781 | J. Lu et al. (2023) |
| 2012ht | 87 | 4/1/13 | 56383.14 | 1014.4 | HD 96781 | J. Lu et al. (2023) |
| 2012ht | 135 | 5/19/13 | 56431.04 | 1902.0 | HD 84834 | This work |
| 2013E | 62 | 3/20/13 | 56371.08 | 507.2 | HD 67526 | J. Lu et al. (2023) |
| 2013E | 67 | 3/25/13 | 56376.08 | 507.2 | HD 77562 | J. Lu et al. (2023) |
| 2013E | 74 | 4/1/13 | 56383.21 | 1014.4 | HD 89662 | J. Lu et al. (2023) |
| 2013aa | 87 | 5/19/13 | 56431.25 | 507.2 | HD 125062 | J. Lu et al. (2023) |
| 2013aa | 100 | 6/1/13 | 56444.22 | 634.0 | HIP 72900 | This work |
| 2013aa | 114 | 6/15/13 | 56458.24 | 507.2 | HD 131057 | This work |
| 2013aa | 145 | 7/16/13 | 56489.17 | 2219.0 | HD 134240 | This work |
| 2013aa | 368 | 2/22/14 | 56710.37 | 2028.8 | HD 127607 | S. Kumar et al. (2023) |

**Table 1**
(Continued)

| SN | Phase (days) | UT Date (MM/DD/YY) | MJD | Exp Time (s) | Telluric Standard | References |
|---|---|---|---|---|---|---|
| 2013aa | 428 | 4/23/14 | 56770.33 | 3550.4 | HD 140425 | S. Kumar et al. (2023) |
| 2013aa | 505 | 7/10/14 | 56848.13 | 4057.6 | HD 140425 | S. Kumar et al. (2023) |
| 2013aj | 70 | 5/19/13 | 56431.21 | 951.0 | HD 125062 | J. Lu et al. (2023) |
| 2013cs | 52 | 7/16/13 | 56489.05 | 1268.0 | HD 115190 | J. Lu et al. (2023) |
| 2013ct | 78 | 7/16/13 | 56489.33 | 1268.0 | HIP 116886 | J. Lu et al. (2023) |
| 2013ct | 179 | 10/25/13 | 56590.15 | 1268.0 | HIP 23691 | This work |
| 2013ct | 205 | 11/20/13 | 56616.15 | 1902.0 | HD 8325 | This work |
| 2013ct | 235 | 12/20/13 | 56646.09 | 1014.4 | HD 4329 | This work |
| 2013fz | 51 | 12/27/13 | 56653.23 | 1521.6 | HD 30252 | J. Lu et al. (2023) |
| 2013fz | 56 | 1/1/14 | 56658.26 | 1014.4 | HD 30252 | J. Lu et al. (2023) |
| 2013gy | 54 | 2/15/14 | 56703.12 | 1268.0 | HD 13281 | J. Lu et al. (2023) |
| 2013gy | 61 | 2/22/14 | 56710.08 | 1014.4 | HD 13281 | J. Lu et al. (2023) |
| 2013gy | 66 | 2/27/14 | 56715.05 | 1014.4 | HD 243374 | J. Lu et al. (2023) |
| 2013gy | 85 | 3/18/14 | 56734.02 | 1268.0 | HD 24083 | J. Lu et al. (2023) |
| 2015F | 68 | 6/1/15 | 57174.96 | 1014.4 | HD 56908 | J. Lu et al. (2023) |
| 2015bp | 62 | 6/1/15 | 57174.21 | 1268.0 | HD 125062 | S. D. Wyatt et al. (2021) |
| LSQ15aae | 59 | 6/2/15 | 57175.27 | 2536.0 | HD 142944 | J. Lu et al. (2023) |
| PS15sv | 60 | 6/1/15 | 57174.29 | 1521.6 | HD 144980 | J. Lu et al. (2023) |

**Note.** SN 2013cs is also known as LSQ13aiz (E. S. Walker et al. 2013), and SN 2015bp is also known as SNhunt281 (S. W. Jha et al. 2015). Nearly half of the spectra are at phases past +100 days, including 31 previously unpublished spectra. The consistent application of telluric corrections is particularly important for this analysis; therefore, we list the telluric standard associated with each spectroscopic observation. The exposure time is the total on-target exposure time, and the phase is in the rest frame with respect to the *B*-band maximum. All spectra listed in this table are presented in Figure 1.

challenging. Single ion PHOENIX/ID synthetic spectra published by B. Friesen et al. (2014) show contributions to the 1.9 $\mu$m emission feature from [Ni II]; additional synthetic spectra produced without including any forbidden lines show a complete lack of spectral features in the 1.9 $\mu$m region, thus implying this feature is mostly composed of forbidden emission lines of Fe, Ni, and Co.

Models and synthetic spectra of SNe Ia in the NIR indicate the strongest Fe and Co lines in the region are [Fe II] $\lambda\lambda$1.967, 2.0027, and 2.047 $\mu$m and [Co III] $\lambda$ 2.0028 $\mu$m (A. Flörs et al. 2020; P. Hoeflich et al. 2021). Although we make no assumptions or line identifications prior to the fitting process, these are the primary neighboring emission lines that will be considered and discussed in addition to [Ni II] $\lambda$1.939 $\mu$m.

At late times, the 1.9 $\mu$m region contains a broad emission feature whose line profile shape evolves over time. In our sample, most transitional phase NIR spectra show a broad line profile with a "shoulder" on the blue side of the emission feature. Several months later, during the nebular phase, the 1.9 $\mu$m feature begins to take on a double-peaked profile whose two peaks become more distinct over time.

The NIR spectra in this sample span phases of +50 to +505 days past peak brightness, so the fit method must be suitable for the line profile shape at both transitional and nebular phases. Therefore, we fit the region with two Gaussian profiles: one on the red side to capture the broad profile that dominates the feature as early as the transitional phase, and the other corresponding to the blue-side emission line that begins as a "shoulder" but evolves into a distinct narrow line profile at nebular phases.

The following sections describe the method used to obtain line velocity and line width measurements for the [Ni II] $\lambda$1.939 $\mu$m line, which is attributed to the blue-side Gaussian. We also define a detection parameter, the Gaussian Peak Ratio ($r$), that measures the relative strength of the NIR Ni feature. Measurement results are reported in Section 4 before the subsequent discussion in the following section.

### 3.1. Fit Method

Emission line measurements were obtained using two Gaussian line profiles plus a flat continuum that was fit using a nonlinear least-squares method with the Python package LMFIT (M. Newville et al. 2014). A flat continuum was used as opposed to a nonzero slope continuum because emission in this region may be a combination of allowed and forbidden lines (G. H. Marion et al. 2009), especially at earlier epochs. We are interested in the evolution of the observed peaks in the emission feature, so using a flat continuum to remove contributions from weaker lines helps further isolate the [Ni II] emission. Two Gaussian line profiles are used to compare the emission between the red and blue peaks of the emission feature, and to determine how much of the total feature could possibly be attributed to the [Ni II] $\lambda$1.939 $\mu$m line.

The red-side edge of the feature is heavily blended at the beginning of the transitional phase and does not provide a usable boundary for the fit region; therefore, a minimum phase of +50 days post peak brightness was chosen for this dataset. The phase cut at +50 days also ensures the SNe are well into the transitional phase and allows for comparison to both the transitional phase and the nebular phase model predictions.

Since high-density burning to produce stable IGEs is expected to occur in the innermost region, the line center of the blue-side Gaussian was then restricted to be within $\pm$0.025 $\mu$m of the rest wavelength. This corresponds to a line velocity range of approximately $\pm$4000 km s$^{-1}$, which is still significantly larger than what is predicted by most explosion models for stable IGEs in the central region. This condition was imposed to ensure distinction between the red and blue halves of the emission feature to see if the blue side contains an emission line that can be detected above the continuum. No restrictions were placed on the central wavelength of the red-side Gaussian. The velocity errors are determined using the uncertainties in measured central wavelength obtained from the fitting process, which is then added in quadrature to the velocity error from the resolution of the instrument in the 1.9 $\mu$m region.

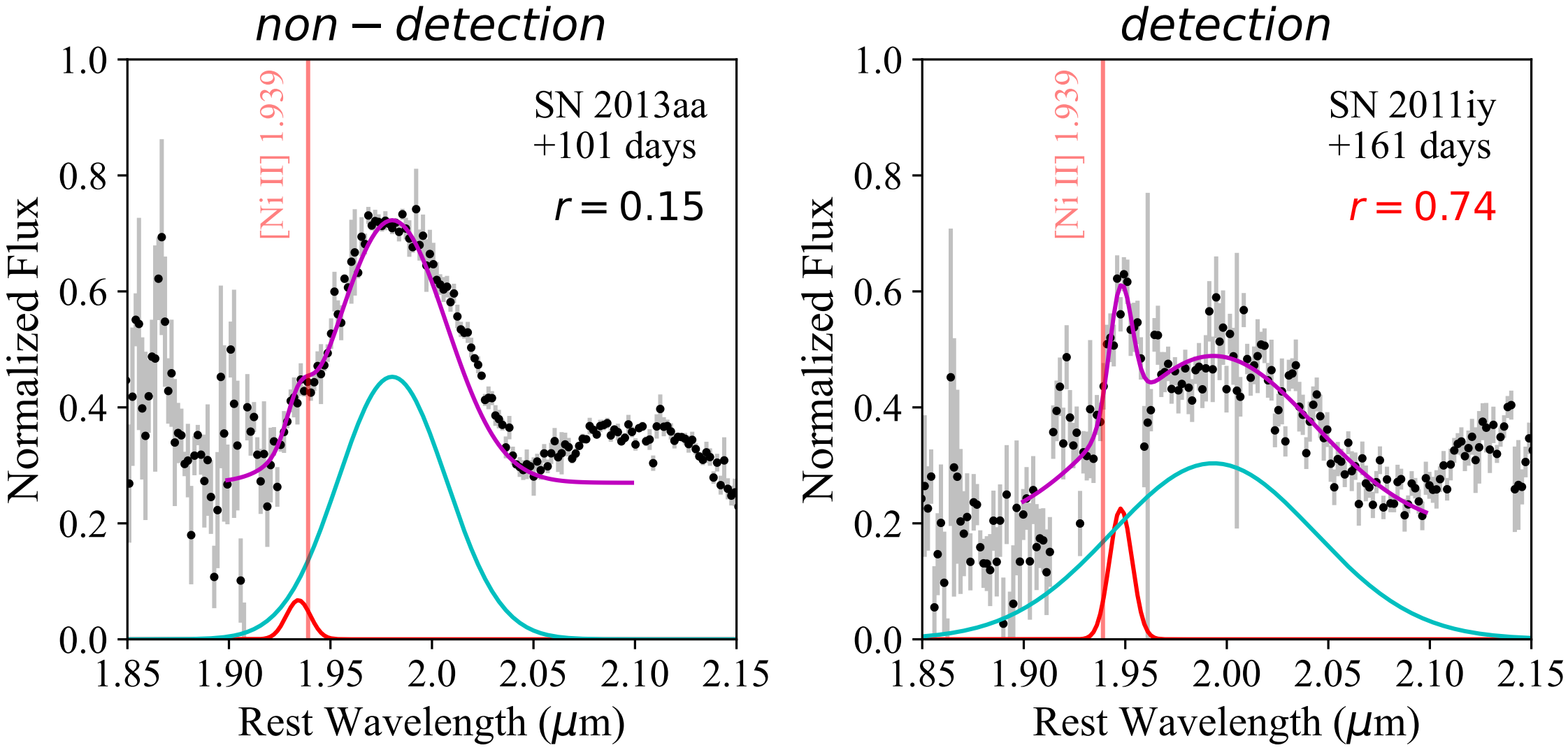

**Figure 2.** Here, we show examples of a nondetection and detection of stable Ni as determined by $r$, the Gaussian Peak Ratio. As discussed in Section 4.1, the cyan Gaussian corresponds to a [Co III] $\lambda$ 2.0028 $\mu$m line and the red Gaussian is the [Ni II] 1.939 $\mu$m emission line. Note that a "nondetection" does not mean the SN has no stable Ni, but the NIR Ni feature is not sufficiently isolated to directly measure the line width and velocity. The detection threshold of $r \geqslant 0.5$ is also used to constrain the phase at which this NIR Ni feature emerges in the SNe Ia in our sample. Further details on the fit method and detection parameters can be found in Section 3.

The best fit for each spectrum was determined using the reduced $\chi^2$ value and visually verifying that the continuum level does not cut into the blue side of the feature, removing contributions from the [Ni II] line. A S/N threshold of S/N $\geqslant$ 5 between 1.98 and 2.05 $\mu$m was chosen for the final sample based on the S/N of spectra that yielded failed fits. The final sample that meets the selection criteria of S/N $\geqslant$ 5 in the 1.9 $\mu$m region and phase of at least +50 days results in 58 NIR spectra of 18 SNe Ia at redshifts ranging from $z = 0.003$ to $z = 0.02$, with a median redshift of $z = 0.008$. These high-S/N spectra and the associated multi-Gaussian fits are plotted in the Appendix.

### 3.2. Detection Method: Gaussian Peak Ratio

We define the Gaussian Peak Ratio ($r$) as the ratio of the measured peak flux of the blue-side Gaussian to the red-side Gaussian. Since the blue side of this feature is attributed to the [Ni II] 1.939 $\mu$m line, the detection threshold was chosen to be $r \geqslant 0.5$ to confirm the presence of a stable Ni feature. This detection threshold also ensures the measurements of the [Ni II] 1.939 $\mu$m line are less susceptible to fit degeneracy.

Examples of a stable Ni detection and nondetection are shown in Figure 2. It is important to note that the NIR [Ni II] line may still be present in spectra with $r \leqslant 0.5$, but this detection criterion ensures the [Ni II] line is sufficiently isolated to obtain line velocity and line width measurements that are unlikely to be affected by significant line blending.

The Gaussian Peak Ratio ($r$) is plotted over time in Figure 3 for all NIR spectra in our sample that meet the subsample criteria described in the previous section. The gray shaded region corresponds to nondetections, most of which occur at phases between +50 and +125 days. The [Ni II] 1.939 $\mu$m line velocity, line width, and Gaussian Peak Ratio measurements are listed in Table 2 for all stable Ni detections in our sample.

## 4. [Ni II] $\lambda$1.939 $\mu$m Line Measurement Results

Based on Figure 3, we detect stable Ni in the NIR in at least 8 out of the 18 SNe Ia in our high-S/N sample. These detections correspond to spectra with a Gaussian Peak Ratio $r \geqslant 0.5$, which means the NIR [Ni II] 1.939 $\mu$m line is sufficiently present to measure the line's peak velocity and gauge its width. Many SNe with "nondetections" in the NIR may still show stable Ni at much later times in the optical (for example, SN 2013aa and SN 2015F), but are considered "nondetections" here because our NIR spectral coverage is limited to earlier epochs. By measuring the Gaussian Peak ratio for all spectra in our sample, we can track the timescale for the emergence of this NIR [Ni II] emission line.

Stable Ni is detected in multiple SNe in our sample at both transitional and nebular phases. The average stable Ni line velocity is $\sim$1200 km s$^{-1}$ and the average measured line width does not exceed $\sim$3500 km s$^{-1}$ for all spectra in this sample. The low velocity indicates most of the stable Ni is located in the innermost region of the SN, while the narrow width confines the emitting region to be within the inner $\pm$3000 km s$^{-1}$ in velocity space.

For most detections in this sample, the [Ni II] 1.939 $\mu$m line is redshifted. The exceptions are SN 2011iv and SN 2015bp, which clearly exhibit blueshifted NIR Ni features. Although the fit method limited the blue-side Gaussian's central wavelength to $\pm$0.025 $\mu$m of 1.939 $\mu$m, nearly all of the measured central wavelengths fall within $\pm$0.011 $\mu$m of 1.939 $\mu$m($\pm$$\sim$1700 km s$^{-1}$). This indicates that the stable nickel is located at lower velocities that coincide with the innermost region of the SN.

Most notably, SN 2011iv and SN 2015bp have the earliest NIR Ni detections in this sample with Gaussian Peak Ratio of

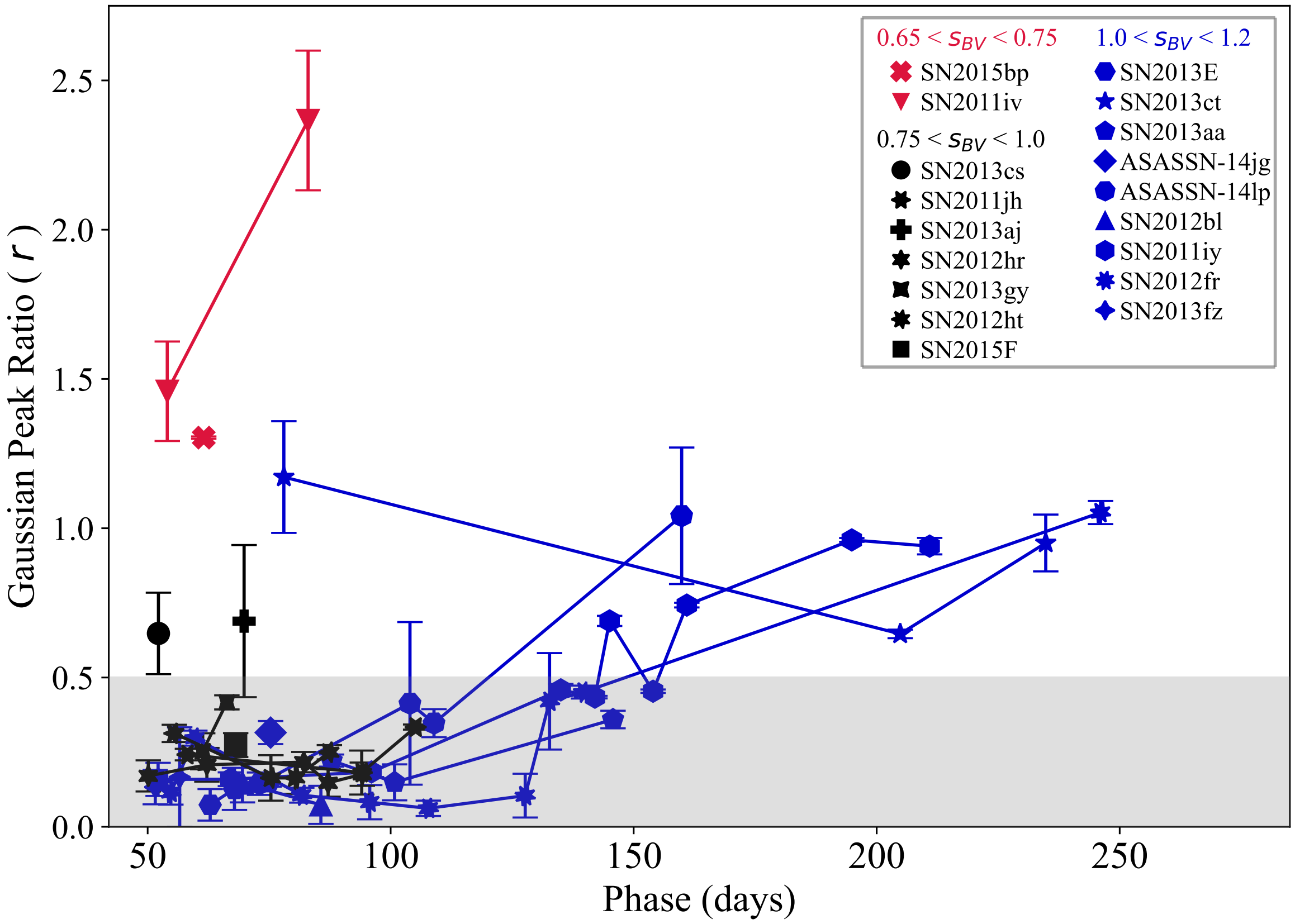

**Figure 3.** The Gaussian Peak Ratio ($r$) over time for all multi-Gaussian fits for the high-S/N sample described in Section 3.1. The ratio is determined by dividing the peak flux of the blue-side Gaussian by the peak flux of the red-side Gaussian. The subluminous 86G-likes with lower $s_{BV}$ values clearly separate from the rest of the sample, and show higher Gaussian Peak Ratios at earlier times, indicating the presence of a stronger NIR [Ni II] $\lambda$1.939 $\mu$m line. Normal-bright SNe Ia also exhibit stable Ni in the NIR, but the [Ni II] $\lambda$1.939 $\mu$m line does not reach our detection threshold until after $\sim$+150 days. Some normal-bright SNe Ia exhibit "detections" at phases before $\sim$+100 days, but these are likely contributions from allowed emission lines that dominate SN Ia spectra at earlier phases. These specific cases of 13ct, 13aj, and 13cs are discussed in Section 4.2.

**Table 2**
Line Measurement Results for All Stable Nickel Detections in This Sample

| SN | Phase (days) | $z_{\rm hel}$ | $s_{BV}$ | Gaussian Peak Ratio ($r$) | Measured Wavelength ($\mu$m) | [Ni II] $\lambda$1.939 $\mu$m Velocity (km s$^{-1}$) | [Ni II] $\lambda$1.939 $\mu$m Width (km s$^{-1}$) |
|---|---|---|---|---|---|---|---|
| 2013cs* | 52 | 0.0092 | 0.964 | 0.65 ± 0.14 | 1.9584 ± 0.0077 | 2980 ± 1170 | 3550 ± 1850 |
| 2015bp | 62 | 0.0041 | 0.692 | 1.30 ± 0.01 | 1.9335 ± 0.0006 | −850 ± 130 | 2840 ± 280 |
| 2013aj | 70 | 0.0091 | 0.800 | 0.69 ± 0.25 | 1.9479 ± 0.0087 | 1380 ± 1350 | 4520 ± 2850 |
| 2011iv | 54 | 0.0065 | 0.701 | 1.46 ± 0.16 | 1.9307 ± 0.0052 | −1290 ± 810 | 5150 ± 1090 |
| 2011iv | 83 | 0.0065 | 0.701 | 2.37 ± 0.23 | 1.9296 ± 0.0024 | −1460 ± 380 | 4630 ± 790 |
| 2013ct* | 78 | 0.0038 | 1.002 | 1.17 ± 0.19 | 1.9640 ± 0.0059 | 3840 ± 900 | 4760 ± 1300 |
| 2013ct | 205 | 0.0038 | 1.002 | 0.65 ± 0.01 | 1.9328 ± 0.0031 | −950 ± 490 | 4590 ± 930 |
| 2013ct* | 235 | 0.0038 | 1.002 | 0.95 ± 0.09 | 1.9596 ± 0.0225 | 3170 ± 3440 | 15300 ± 10010 |
| ASAS14lp | 160 | 0.0051 | 1.027 | 1.04 ± 0.23 | 1.9402 ± 0.0047 | 180 ± 720 | 2410 ± 2030 |
| 2011iy | 145 | 0.0043 | 1.033 | 0.69 ± 0.02 | 1.9476 ± 0.0022 | 1320 ± 350 | 3250 ± 1060 |
| 2011iy | 161 | 0.0043 | 1.033 | 0.74 ± 0.01 | 1.9478 ± 0.0008 | 1360 ± 160 | 1720 ± 300 |
| 2011iy | 195 | 0.0043 | 1.033 | 0.96 ± 0.01 | 1.9475 ± 0.0009 | 1310 ± 170 | 2470 ± 370 |
| 2011iy | 211 | 0.0043 | 1.033 | 0.94 ± 0.03 | 1.9468 ± 0.0018 | 1200 ± 280 | 2250 ± 700 |
| 2012fr | 246 | 0.0054 | 1.107 | 1.05 ± 0.04 | 1.9486 ± 0.0018 | 1480 ± 290 | 2060 ± 590 |

**Note.** The detection criteria are a Gaussian Peak Ratio $r \geqslant 0.5$, as described in Section 3.2. The line velocities and widths are calculated using the fit results of the blue-side Gaussian from the multi-Gaussian fit. Rows marked with an asterisk identify spectra with possible contributions from allowed emission lines and are not included in the average velocity and line width values computed for this sample. These exceptions are detailed in Section 4.2.

$r \geqslant 1.0$. These higher Gaussian Peak Ratio values also reflect how prominent the [Ni II] 1.939 $\mu$m line is in these SNe. SN 2011iv and SN 2015bp are the two least luminous SNe Ia in this sample and belong to the "1986G-like" subclass. 86G-likes have $\Delta m_{15}(B)$ and peak $B$-band magnitude values that fall at the fainter end of the normal SNe Ia distribution and are thus a "transition" between underluminous 91bg-likes and "normal-bright" SNe Ia. Interestingly, 86G-likes are subluminous only in optical bands and have similar NIR peak magnitudes when compared to normal SNe Ia (K. Krisciunas et al. 2009).

Although the two subluminous 86G-like objects in the sample show a difference in velocity shift direction, the measured NIR Ni widths are not dissimilar to the widths measured in normal-bright SNe Ia. SN 2015bp has a narrower NIR Ni feature ($\sim$3000 km s$^{-1}$) than SN 2011iv ($\sim$4000–5000 km s$^{-1}$), but the widths are comparable to NIR Ni widths for all other detections in this sample. In our sample, both SN 2011iv and SN 2015bp have limited NIR spectra at late times due to the fact that they became fainter much faster than other SNe Ia, but nebular-phase optical spectra are available in the literature for comparison.

### 4.1. Line Identification: [Co III] $\lambda$2.0028 $\mu$m

As shown in Figure A4, the peak flux of the right-side Gaussian (shown in cyan) consistently decreases for all SNe in the sample with NIR spectra past +100 days. This coincides with the timescale of $^{56}$Co decay (S. A. Colgate & C. McKee 1969) and may be evidence that the right side of the feature is dominated by Co emission.

The time evolution of the central wavelength of the right-side Gaussian tends to remain constant or redshift slightly over time for most objects in this sample. For the spectra that do exhibit a redshift in the right-side Gaussian, it evolves toward the rest wavelength of [Co III] $\lambda$2.0028 $\mu$m for all nebular-phase spectra in the sample, suggesting the red side of the feature is likely composed of this Co feature, which is predicted to be the strongest Co line in this wavelength region (A. Flörs et al. 2020; P. Hoeflich et al. 2021).

Since the Gaussian Peak Ratio ($r$) is defined as the peak flux of the left-side Gaussian divided by the peak flux of the right-side Gaussian, the increase in $r$ over time may be due to the decreasing strength of the right-side Gaussian instead of an increasing strength of the left-side Gaussian. If the right-side Gaussian is indeed [Co III] $\lambda$2.0028 $\mu$m , this Co feature will weaken as Co decays, and by comparison, the neighboring [Ni II] will appear stronger over time as the innermost and higher-density regions of the SN ejecta are revealed.

The left panel of Figure 4 shows the measured peak of both the Ni and Co emission lines, which exhibit a similar decline between $\sim$+100 and $\sim$+250 days. This follows a similar timescale to the overall light curve of SNe Ia at late times, with the SN becoming fainter over time and the ejecta becoming more optically thin. The peaks of the Ni and Co lines reach similar strengths over time, especially after $\sim$+150 days. This is the time frame where the Gaussian Peak Ratio($r$) for normal-bright SNe Ia emerges past the detection threshold ($r \geqslant 0.5$) and indicates that the emission feature is not dominated by the behavior of a single line at these later times.

Another possibility is several strong [Fe II] lines near the 1.9 $\mu$m region, specifically [Fe II] $\lambda\lambda$1.967, 2.0027, and 2.047 $\mu$m (P. Hoeflich et al. 2021). At first glance, the central wavelength appears to evolve toward the rest wavelength of [Fe II] $\lambda$2.007 $\mu$m. However, the [Fe II] $\lambda$2.007 $\mu$m and $\lambda$2.047 $\mu$m lines are expected to have similar strengths (P. Hoeflich et al. 2021) and should both be detected if present. As these are both spectral lines of the same ion and the emission should originate from the same physical region, we would expect to see both of these strong [Fe II] lines at similar velocities. As mentioned in Section 3.1, the continuum level is chosen based on the local minima of the emission feature, which occurs on the red side of the feature in the majority of the sample. The lack of flux on the red side of the feature shows no prominent blueshifted [Fe II] $\lambda$2.047 $\mu$m line and therefore the presence of the [Fe II] $\lambda$2.007 $\mu$m is unlikely as well. Thus, we identify the right-side Gaussian as being composed mostly of [Co III] $\lambda$ 2.0028 $\mu$m for the majority of spectra in this sample.

The two-Gaussian method fits the 1.9 $\mu$m region well for the majority of the spectra in this sample. However, for spectra past $\sim$+200 days post peak brightness, this region may be better fit with three or more line profiles. This is consistent with the timescales of $^{56}$Co decay, which would diminish the [Co III] $\lambda$2.0028 $\mu$m line while nearby [Fe II] lines (P. Hoeflich et al. 2021) become more prominent over time. Significant amounts of Co remain around $\sim$+200 days (M. J. Childress et al. 2015), so it is not surprising that strong lines of both Co and Fe would be present at this time.

The emergence of an [Fe II] feature in the 1.9 $\mu$m region at late times does not challenge the detection of the narrow [Ni II] $\lambda$1.939 $\mu$m line because the [Ni II] feature is consistently detected at much earlier times. This can be clearly seen in the seven spectra of SN 2011iy observed between +135 and +211 days.

One final possibility is that the 1.9 $\mu$m feature contains no nickel and is a blend of iron and cobalt lines; however, the measured emission line parameters and their subsequent evolution over time show the strongest contributions to this emission feature are likely from [Ni II] $\lambda$1.939 $\mu$m and [Co III] $\lambda$2.0028 $\mu$m. This is further corroborated by the JWST observations discussed in Section 5.3.

### 4.2. Time Evolution

The timing of the NIR Ni detections yields further interesting results. As shown in Figure 3, the earliest Ni detections in this sample occur in the subluminous 86G-like SNe Ia at transitional phases. These are the strongest stable Ni detections in this sample with the highest Gaussian Peak Ratio ($r$) values. Several normal-bright SNe Ia in our sample also show stable nickel in the NIR at relatively early times.

As mentioned in Section 2.2, the telluric corrections method for ground-based NIR observations from W. D. Vacca et al. (2003) uses the optical $B$- and $V$-band magnitudes of the telluric standard star to determine the flux calibration for the observed NIR spectrum. Further testing of the accuracy of the flux calibration in the telluric regions requires additional observations that are beyond the scope of this work, but we hope this dataset can provide informative estimates for future NIR studies. Thus, we have opted to report approximate absolute line luminosities with this important caveat in mind.

In order to maintain a consistent method across the entire dataset, we use the heliocentric redshift reported in S. A. Uddin et al. (2024), assume a standard $\Lambda$CDM cosmology ($H_0 = 72$ km s$^{-1}$ Mpc$^{-1}$, $\Omega_m = 0.27$, $\Omega_\Lambda = 0.73$), and ignore extinction to calculate the distances for all SNe in our sample to estimate absolute line luminosities. The right panel of Figure 4 shows the approximate absolute line luminosity for all measured Ni and Co lines in our sample.

SNe 2011iy, 2012fr, and 2013ct are three normal-bright objects with spectra in both the transitional and nebular phases. Early in the transitional phase ($\sim$50–80 days), the double-peaked morphology of the 1.9 $\mu$m region is evident in all three SNe, but the blue-side emission line is almost merged with the red side, showing a blended profile with no isolated peaks. If the blue-side emission line is interpreted as [Ni II] 1.939 $\mu$m, this feature is redshifted by as much as 4000 km s$^{-1}$. The lack of isolated peaks also results in large uncertainties in the measurements of these features.

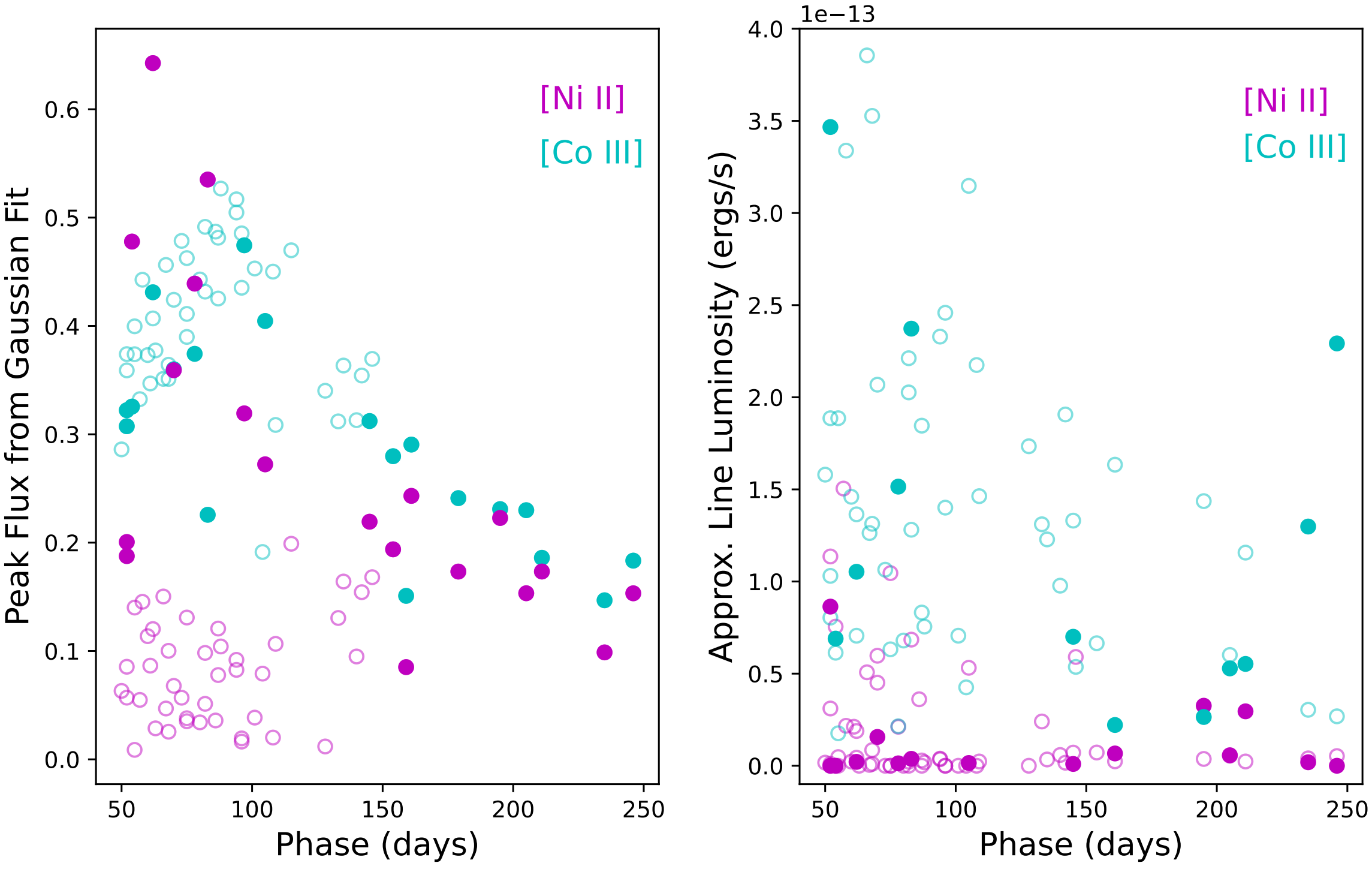

**Figure 4.** The measurement results of each individual line profile from the fit method described in Section 3.1 are shown over time for all NIR spectra in this sample, including both detections (filled circles) and nondetections (open circles) of stable nickel. Left: The measured peak flux of each individual Gaussian profile is shown over time for all non-null fit results as described in Section 3.1. The reduced number of nondetections after ∼+150 days is primarily due to measured peak Ni flux values that are smaller than the observed flux error or a failed fit due to low S/N. Both the Ni and Co lines show a similar decline over time as the SNe become fainter and more optically thin, indicating that the behavior of the Gaussian Peak Ratio ($r$) shown in Figure 3 is not solely due to one line dominating the feature at later times. Right: The approximate absolute line luminosity of each measured Ni and Co line profile is shown over time. This is an approximate luminosity because the flux calibration in telluric regions relies on extrapolation from optical observations. Unlike the left panel of this figure, spectra with nearly no flux in the Ni line (i.e., measured peak line flux is less than observed flux error) are included in the right panel to show the approximate line luminosity of the [Co III] feature, which is still often present even when stable Ni is not conclusively detected. Further discussion on the individual line luminosities and important caveats is detailed in Sections 2.2 and 4.2.

Later in the transitional phase (∼+80–100 days), the blue-side feature is observed to disappear completely in SNe 2011iy and 2012fr, forming a single feature in the region. Then, during the nebular phase (⩾+100 days), another blue-side feature emerges with isolated peaks that more closely match the rest wavelength of [Ni II] 1.939 $\mu$m. This suggests that the feature that emerges early in the transitional phase and the one that emerges in the nebular phase may not be of the same line, with the former having possible contributions from Fe II 1.967 $\mu$m or other permitted IGE lines. Note that the merged line profile shape is common in the earlier spectra of normal-bright SNe Ia in our sample. Only two of these spectra crossed our detection threshold: 2013cs at +52 days and 2013ct at +78 days. They have been noted in figures and tables as detections that are not likely to be attributed to stable $^{58}$Ni.

SN 2013cs, SN 2013aj, and SN 2013ct are all normal-bright SNe Ia with a Gaussian Peak Ratio $r \geqslant 0.5$ at phases ⩽+100 days. SN 2013cs shows the earliest NIR nickel detection in this sample at +52 days. The presence of stable nickel in this SN is corroborated by multiple nebular-phase optical spectra with [Ni II] 7378 Å features (M. L. Graham et al. 2017; K. Maguire et al. 2018).

SN 2013aj is an interesting case because it has an $s_{BV}$ value of 0.8, which is low compared to the $s_{BV}$ values of other normal-bright SNe Ia in this sample, but is still a higher $s_{BV}$ value than the two subluminous 86G-like SNe. SN 2013aj exhibits NIR [Ni II] at +70 days, which is at a similar phase to the stable Ni detections in the two 86G-like objects. However, the Gaussian Peak Ratio of SN 2013aj is much lower, indicating a weaker Ni feature compared to the 86G-likes. In this regard, SN 2013aj may be showing characteristics that bridge the gap between 86G-likes and normal SNe Ia.

SN 2013ct has NIR Ni detections at both transitional and nebular phases, but the morphology of the measured Ni line shows an unusual time evolution when compared to other objects in this sample. The first Ni detection in SN 2013ct occurs at +78 days with $r = 1.17$, but nebular-phase spectra at +205 days and +235 days have $r \leqslant 1.0$. SN 2013ct meets the detection criteria of a Gaussian Peak Ratio $r \geqslant 0.5$, but the broad and flat profile of the blue side of the 1.9 $\mu$m feature results in a larger velocity error than other detections in this sample. The time evolution of the left-side Gaussian in SN 2013ct is notably different than other SNe Ia with spectra at similar phases.

At these earlier phases, the presence of allowed lines may still be contributing to the 1.9 $\mu$m feature, and the strength of the neighboring Co feature could potentially shift the blue-side Gaussian's central wavelength redward. Furthermore, the SN ejecta is not completely optically thin at these earlier phases, and recent studies indicate a small pseudo-photosphere may linger in some SNe Ia well into the nebular phase (P. Hoeflich et al. 2021; C. Ashall et al. 2024; J. M. DerKacy et al. 2024). As detailed in Section 4.1, the nearby Co feature diminishes over time, so this may be part of the reason why the largest redshifts measured in this sample occur in some of the earliest spectra.

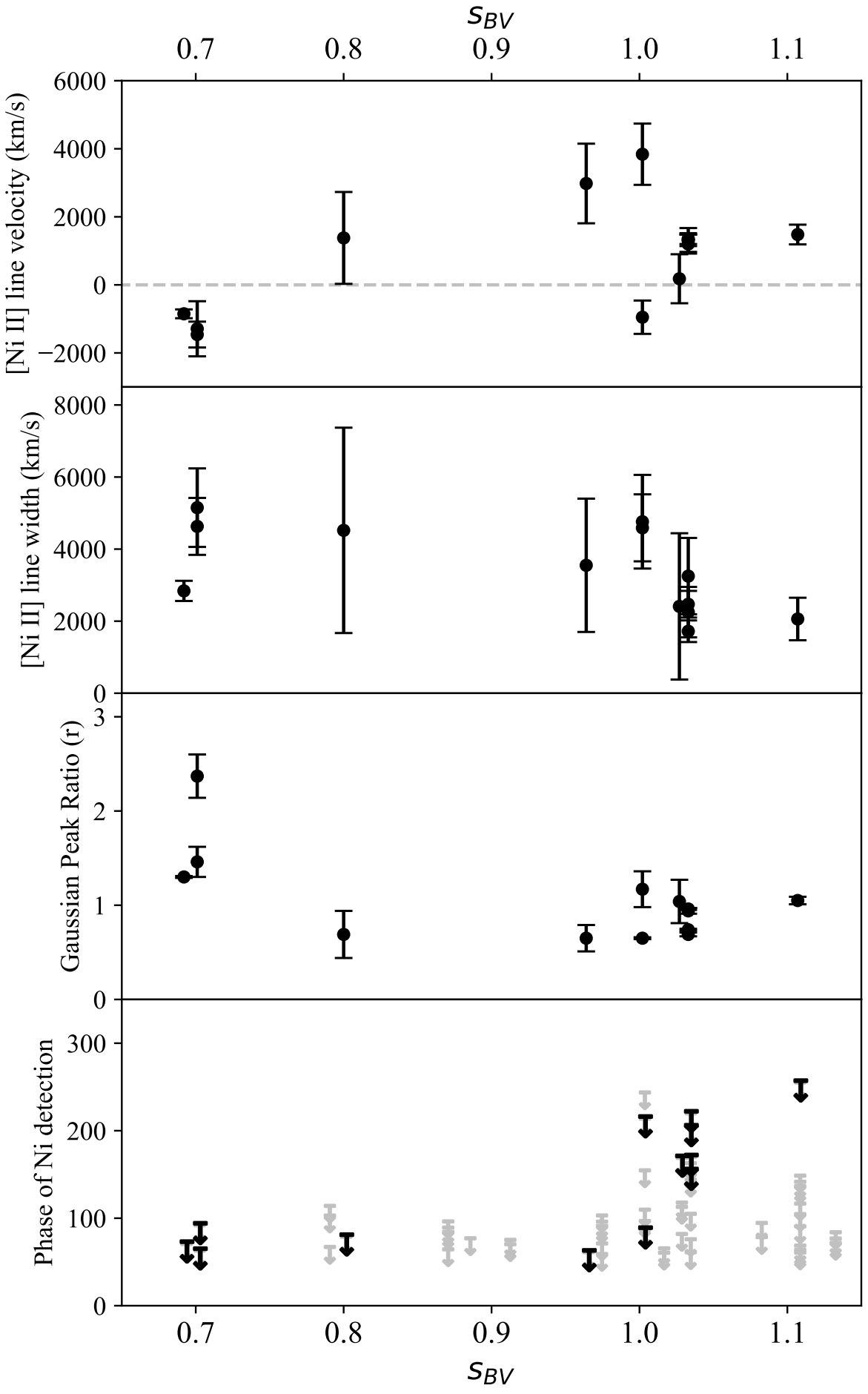

**Figure 5.** All measured parameters of the NIR [Ni II] $\lambda$1.939 $\mu$m feature listed in Table 2 are plotted with respect to $s_{BV}$, the color-stretch parameter. The lower panel shows the phase of each Ni detection in our sample (in black) with respect to peak brightness in the $B$ band, providing upper limits for the onset of stable Ni emission features for the SNe Ia in our sample. Gray points correspond to nondetections within the high-S/N subsample.

As shown in Table 2 and Figure 5, approximately half of the stable Ni detections in this sample occur at phases past +100 days. SN 2011iy and SN 2012fr have the most robust time-series spectroscopy in this sample, each with 10 NIR spectra with an S/N $\geqslant$ 5 in the 1.9 $\mu$m region. The narrow profile of the [Ni II] $\lambda$1.939 $\mu$m line emerges ∼+140 days and continues to be detected in both SNe at phases past +200 days.

Furthermore, six normal-bright SNe Ia in this sample have redshifted NIR [Ni II] lines. Of the SNe with multiple stable Ni detections over time, most seem to show little to no change in their measured [Ni II] $\lambda$1.939 $\mu$m line velocities. For example, SN 2011iy maintains a consistent [Ni II] velocity of ∼1300 km s$^{-1}$ between +145 and +211 days.

One possible consideration of redshifted nebular-phase emission lines is asymmetries in the SN ejecta along our line of sight (K. Maeda et al. 2010). Depending on the geometry of the explosion, asymmetric distributions of material in the ejecta can affect the shape of emission line profiles (P. Hoeflich et al. 2021; S. Bose et al. 2025). If the consistent redshift of nebular-phase NIR [Ni II] lines is caused only by asymmetries along our line of sight, that implies the geometry of the ejecta is identical for at least six normal-bright SNe Ia.

The possibility of these six well-observed explosions having identical geometry while exhibiting other different observed properties (i.e., decline rate, peak brightness, color evolution, variation in other spectral features) is incompatible with our current understanding of SN Ia physics. Nevertheless, the narrow NIR [Ni II] line is well-fit by a Gaussian line profile in our sample, implying a symmetric distribution of Ni confined to the low-velocity inner region of the SN ejecta. We therefore consider it unlikely that redshifted Ni lines are purely due to viewing angle effects in six different well-observed SNe Ia.

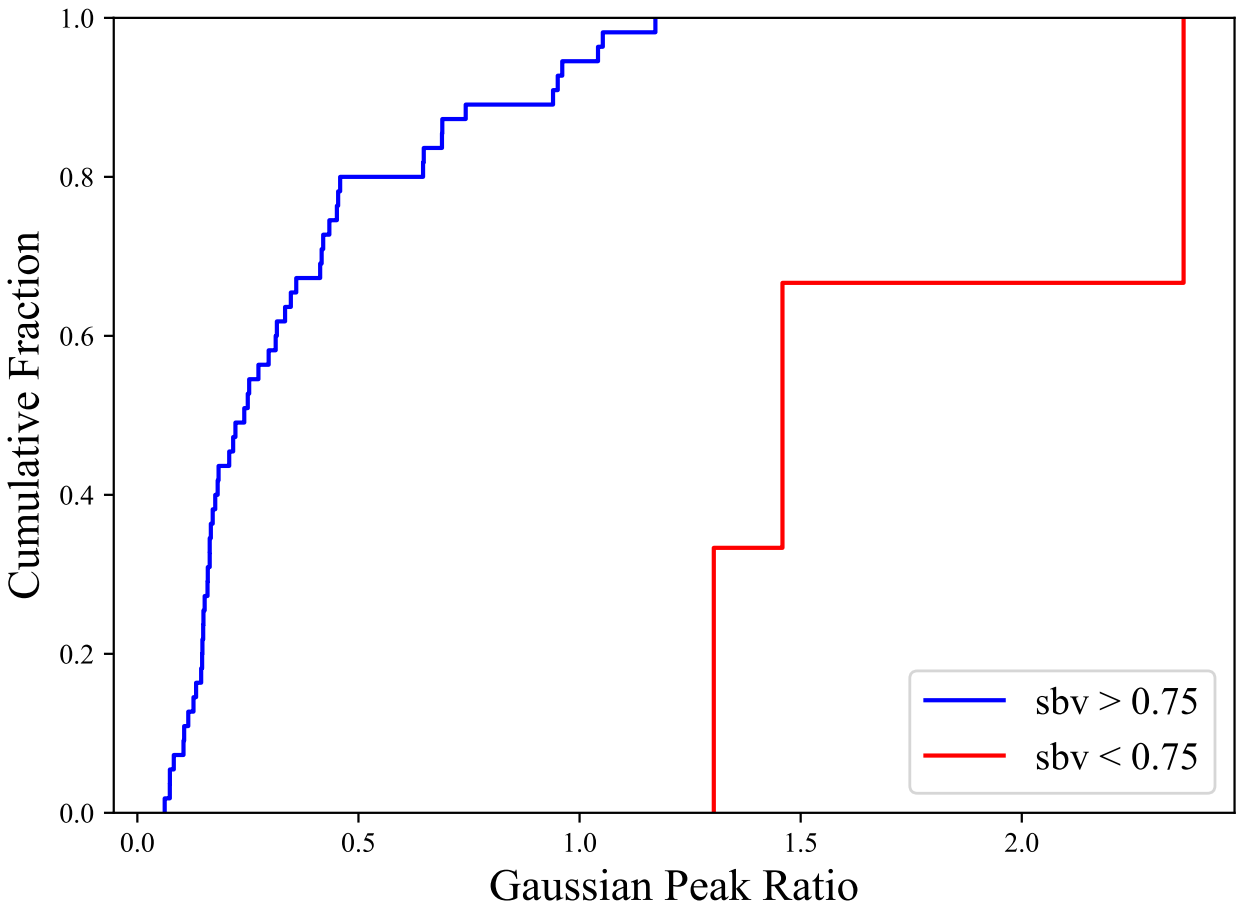

**Figure 6.** Cumulative distribution plots resulting from a two-sample KS test as described in Section 4.3. The blue curve includes 55 spectra of 16 normal-bright objects in our sample, whereas the red curve comprises 3 spectra of the 2 subluminous 86G-like objects. The results of the two-sample KS test imply the two groups of SNe (distinguished by their $s_{BV}$ values) are unlikely to come from the same underlying probability distribution, and therefore the red and blue curves above may be describing two separate populations. However, we caution that the small sample size of the $s_{BV} \leqslant 0.75$ group limits further interpretation of this statistical result, and future observations may reveal the full range of Gaussian Peak Ratio values from subluminous 86G-like SNe Ia.

### *4.3. KS Test*

The two-sample Kolmogorov–Smirnov test (KS test) is a way to test the likelihood of two samples coming from the same underlying probability distribution. Here, the two samples we test are individual spectra sorted into two groups based on $s_{BV}$ values. All spectra with an S/N $\geqslant$ 5 in the 1.9 $\mu$m region are included in the two groups, resulting in 55 spectra in the $s_{BV} \geqslant 0.75$ group and 3 spectra in the $s_{BV} \leqslant 0.75$ group. This separates the subluminous 86G-like objects from the normal-bright SNe Ia regardless of the phase of each spectrum. Figure 6 shows the cumulative distribution plots versus the Gaussian Peak Ratio for the two groups. We assume a normal probability distribution for both groups and place no restrictions on the range of possible Gaussian Peak Ratio values.

The two-sample KS test results in a p-value of $6.48 \times 10^{-5}$. This low value rejects the null hypothesis and suggests the two groups do not come from the same underlying probability distribution. The $s_{BV} \leqslant 0.75$ group only contains three spectra, which is not a statistically significant sample size, and therefore, we limit the interpretation of the red curve shown in Figure 6. On the other hand, the $s_{BV} \geqslant 0.75$ group contains 55 spectra that span both transitional and nebular phases and includes spectra both with and without stable Ni detections.

The cumulative distribution plot shows that SNe Ia with $s_{BV}$ values $\geqslant$0.75 span a continuous range of Gaussian Peak Ratio values, but even the highest Gaussian Peak Ratio value for normal-bright SNe Ia is less than the lowest Gaussian Peak

Ratio value for subluminous 86G-likes. This result is independent of phase information, but 16 of the 18 SNe Ia in the high-S/N sample are in the $s_{BV} \geqslant 0.75$ group, and nearly half of the spectra in that group are at phases past +100 days. Therefore, the cumulative distribution plot is not skewed by a lack of data at later phases, nor is it dominated by the evolution of an individual SN.

Regardless of the time evolution of the individual SNe in each $s_{BV}$ group, normal-bright SNe Ia at phases past +200 days still do not have Gaussian Peak Ratio values as large as the subluminous 86G-likes at phases before +100 days. Therefore, we examine the time evolution of subluminous 86G-likes separately in Section 5.5 and use that information to inform the NIR follow-up strategy suggested in Section 6.

## 5. Discussion

The following discussion focuses on how these NIR detections of stable Ni inform our understanding of SN Ia origins. Sections 5.1, 5.2, and 5.3 compare our NIR results to previously published stable Ni detections in other wavelength regimes. Finally, the NIR behavior of the SNe Ia in our sample is explored in the context of various explosion mechanisms in Section 5.4.

### 5.1. Optical [Ni II] λ7378 Å

The NIR Ni detections are further supported by previously published optical results. Nearly all of the SNe Ia in this sample with NIR Ni detections have previously published velocity measurements of the optical [Ni II] λ7378 Å emission line. The optical [Ni II] λ7378 Å line is not detected until much later in the nebular phase (i.e., past +200 days) for most normal SNe Ia. Although the [Ni II] λ7378 Å line is the strongest nebular-phase Ni feature in the optical, it is heavily blended with neighboring emission lines and is often measured using a multi-Gaussian fit.

Six SNe Ia in our sample with NIR Ni detections have previously published optical [Ni II] λ7378 Å line velocity measurements: SN 2012fr (M. L. Graham et al. 2017; K. Maguire et al. 2018), SN 2013cs (M. L. Graham et al. 2017; K. Maguire et al. 2018), SN 2013ct (K. Maguire et al. 2018), SN 2015bp (S. Srivastav et al. 2017), SN 2011iv (P. A. Mazzali et al. 2018), and ASASSN-14lp (B. E. Stahl et al. 2020). These previously published optical [Ni II] measurements are compared to NIR [Ni II] velocities in Figure 7.

Optical [Ni II] velocities of the two 86G-likes in our sample agree with the blueshifts measured using the NIR [Ni II] λ1.939 μm line (S. Srivastav et al. 2017; P. A. Mazzali et al. 2018). Interestingly, the optical [Ni II] λ7378 Å feature is significantly stronger than the neighboring [Fe II] λ7155 Å feature in SN 2011iv (C. Gall et al. 2018; P. A. Mazzali et al. 2018). SN 2015bp also shows a strong, narrow emission feature on the red side of the optical 7200 Å blended feature (S. Srivastav et al. 2017). This is unusual as the majority of nebular-phase optical spectra exhibit either a stronger [Fe II] line or nearly equal strength [Fe II] and [Ni II] lines in the blended double-peaked feature around 7200 Å (K. Maguire et al. 2018; A. Polin et al. 2021; M. L. Graham et al. 2022). The NIR [Ni II] λ1.939 μm feature becomes more prominent between +54 and +83 days in SN 2011iv, so the presence of a strong optical [Ni II] feature approximately 200 days later may be a continuation of this trend.

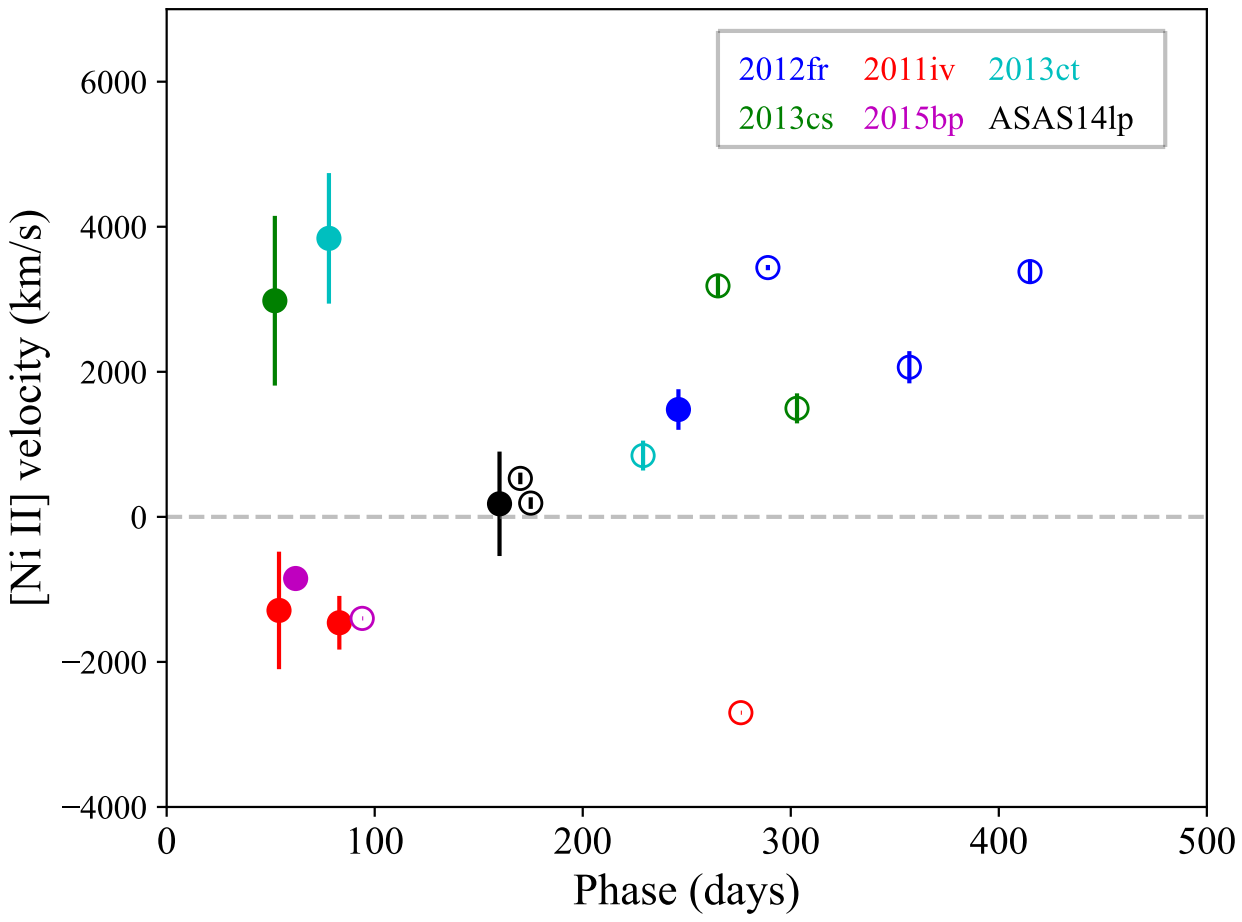

**Figure 7.** Six SNe Ia in our sample have previously published optical [Ni II] 7378 Å line velocities (M. L. Graham et al. 2017; S. Srivastav et al. 2017; K. Maguire et al. 2018; P. A. Mazzali et al. 2018; B. E. Stahl et al. 2020), which are compared here to velocities measured using the NIR [Ni II] 1.939 μm line. Previously published optical [Ni II] velocities were obtained using different fit methods, but the direction of the velocity shift (i.e., redshifted or blueshifted) consistently agrees with measurements made in the NIR. If reported, velocity errors from optical measurements are included as error bars.

Most previously published optical detections of stable Ni are at phases past ∼+200 days, with the notable exception of SN 2015bp, which is one of the subluminous 86G-likes in this sample. The low Ni line velocity ($\leqslant$2000 km s$^{-1}$) measured in both the optical and NIR implies the ejecta has become optically thin enough before +100 days to reveal the innermost region of the SN ejecta. This is consistent with other observational evidence indicating faster-declining SN Ia subtypes may go nebular sooner than normal-bright SNe Ia (K. Krisciunas et al. 2009).

Regardless of the fit method used in previously published optical studies, the direction of the velocity shift (i.e., blueshifted or redshifted) measured using the NIR [Ni II] line is in agreement with measurements from the optical. However, the velocities measured using optical spectra are often larger than those measured in the NIR.

This discrepancy between optical and NIR velocities of the same ion at similar phases has been seen in other SNe Ia. S. Kumar et al. (2023) showed that [Fe II] velocities measured using optical features result in values ∼1000 km s$^{-1}$ larger than [Fe II] velocities measured in the NIR at similar phases. This offset in measured velocity between the optical and NIR is seen for both blueshifted and redshifted [Fe II] lines. Although S. Kumar et al. (2023) focused on the velocities of nebular-phase Fe lines, the same behavior is seen here in the Ni lines of SNe in this sample.

S. Kumar et al. (2023) suggest that this velocity difference is due to a difference in opacity at optical and NIR wavelengths, and the NIR may be probing a physically deeper region than optical spectra at the same phase. After ∼+100 days, we can be reasonably certain that all Ni emission originates from stable material due to the short half-life of radioactive $^{56}$Ni. On the other hand, the main radioactive isotope of Fe in SNe Ia, $^{55}$Fe, has a half-life of ∼1000 days (S. A. Colgate & C. McKee 1969). At late times, the Fe in the ejecta is a combination of radioactive and stable material. If we are finding similar trends in our detection of both stable and radioactive material, this may reflect the physical conditions within the ejecta, though, of course, line blending with neighboring features must still be considered.

### 5.2. NIR [Ni II] 1.939 μm in SN 2014J

SN 2014J has been extensively observed through very late times, and previously published NIR spectra can be a useful comparison to the FIRE spectra used in this study. Its nearby location in M82 has enabled multiwavelength follow-up very late into the nebular phase compared to most other SNe Ia (e.g., C. Ashall et al. 2014; B. Friesen et al. 2014; D. J. Sand et al. 2016; S. Dhawan et al. 2018).

Most previously published detections of the NIR [Ni II] 1.939 $\mu$m feature are from spectra obtained well into the nebular phase; yet some studies predict the presence of this emission line at much earlier times. B. Friesen et al. (2014) compare PHOENIX/1D models to observed NIR spectra of four SNe Ia (including SN 2014J) during the transitional phase when the ejecta is not entirely optically thin. Therefore, the models used in this study test different combinations of allowed and forbidden lines. Both the synthetic spectra and the observed spectra in B. Friesen et al. (2014) exhibit an emission feature in the 1.9 $\mu$m region, but the synthetic spectra consistently produce a highly blueshifted NIR [Ni II] line.

In our sample, the vast majority of [Ni II] 1.939 $\mu$m detections occur at redshifted wavelengths, similar to the observed spectra from B. Friesen et al. (2014). The velocity uncertainties from our NIR Ni line velocities do imply the possibility of slightly blueshifted NIR [Ni II] lines, but none are large enough to match the significant blueshifts produced by the models in B. Friesen et al. (2014).

The synthetic spectra from B. Friesen et al. (2014) also show little migration in the central wavelength of this [Ni II] feature over time. This behavior is consistent with the time evolution of normal-bright SNe with multiple late-time stable Ni detections in our sample, such as SN 2011iy. However, we note the stable Ni detections in SN 2011iy occur at phases ~100 days later than the phase range examined by B. Friesen et al. (2014).

S. Dhawan et al. (2018) published multiple nebular-phase spectra of SN 2014J and identified the broad feature between 1.9 and 2.0 $\mu$m as [Ni II] 1.939 $\mu$m. This feature persists, though the strength decreases over time between +408 and +478 days. S. Dhawan et al. (2018) use a comparison to non-LTE models to obtain emission line parameters and report a NIR [Ni II] line width of ~11,000 km s$^{-1}$ and a redshifted line velocity of ~800 km s$^{-1}$at +478 days. The redshifted [Ni II] peak velocity is consistent with the normal-bright SNe Ia in our sample, but the NIR [Ni II] line width reported in S. Dhawan et al. (2018) is much broader than the line widths measured in our sample. If the emission feature at ~ 1.9 $\mu$m is attributed entirely to stable Ni, this would imply an extended region of stable nickel that extends to velocities much larger than other stable IGEs, as determined by their reported measurements of [Fe II] and [Co II] lines. We note that this study obtained these emission line measurements via comparison to synthetic spectra as opposed to direct fits to the observed spectra and did not include any continuum, but the discrepancy in measured line widths is large enough to not be due to any differences in measurement method.

If we compare to Figure 2 in S. Dhawan et al. (2018), the single broad line profile in SN 2014J at +478 days is remarkably different from the multiply peaked profiles seen in SN 2011iy and SN 2012fr. It is possible this discrepancy may be due to the emergence of other emission features during the ~200 day phase difference between the latest spectrum in our FIRE sample at +246 days and SN 2014J at +478 days. The lower S/N of the SN 2014J observations may also obscure a narrow [Ni II] feature if it is present. Another possibility is an intrinsic difference between SN 2014J and other normal-bright SNe. Nebular-phase [Fe III] lines diminish in both the optical (P. A. Mazzali et al. 2020) and NIR (S. Kumar et al. 2023) earlier in SN 2014J than in other normal-bright SNe Ia, possibly indicating SN 2014J may have cooled at a faster rate than other objects. Further analysis of SN 2014J is beyond the scope of this paper, but future observations of the 1.9 $\mu$m region in other SNe Ia at phases between +300 and +400 days may provide the answer.

### 5.3. Detecting Stable Ni with JWST

The launch of the JWST has enabled spectroscopic observations of SNe Ia in the NIR and MIR that have not been possible since the era of the Spitzer Space Telescope (SST). Space telescope observations are unhindered by telluric features from Earth's atmosphere, so JWST is ideal for observing the 1.9 $\mu$m region in SNe Ia. However, current programs can only observe a limited number of targets, so it will take several years for JWST to observe a larger sample of nebular-phase SNe Ia.

Recent studies (J. M. DerKacy et al. 2023, 2024; L. A. Kwok et al. 2023; C. Ashall et al. 2024) have published nebular-phase JWST spectra of two nearby SNe Ia: SN 2021aefx and SN 2022xkq. These two objects are of particular interest to this work because SN 2021aefx is a normal-bright SN Ia (C. Ashall et al. 2022), whereas SN 2022xkq is an underluminous 91bg-like SN Ia (J. Pearson et al. 2024). We note that SN 2022xkq is a different subtype than SN 2011iv and SN 2015bp, but the lower luminosities of these SNe may hint at similar origins.

Initial nebular-phase JWST observations of SN 2021aefx utilized lower-resolution modes to expand spectral coverage into wavelengths that have been rarely observed (J. M. DerKacy et al. 2023; L. A. Kwok et al. 2023). The JWST/NIRSpec spectrum of SN 2021aefx at +255 days exhibits a NIR [Ni II] 1.939 $\mu$m feature, with a reported line velocity of $300 \pm 1000$ km s$^{-1}$ (L. A. Kwok et al. 2023). This supports our identification of the same feature as Ni in other SNe Ia. However, the relatively low spectral resolution ($R \sim 40$–250) and previously uncertain wavelength calibration make it difficult to constrain the exact velocity, particularly for narrow features. However, the lack of telluric absorption from the atmosphere is still a major advantage for observing the 1.9 $\mu$m region with JWST.

The 1.9 $\mu$m feature in the JWST spectra of 2021aefx shows a double-peaked profile with [Ni II] comprising the blue side of the feature, which is very similar to the normal-bright SNe Ia in our sample. We consider the possibility of imprecise telluric corrections to ground-based observations that may result in spectral features from the atmosphere remaining in the science spectrum, possibly affecting the morphology of emission lines from the supernova. However, the consistent detection of a narrow Ni line in multiple SNe Ia in our sample indicates this feature is real, but additional corroboration from space telescope observations confirms the structure of the 1.9 $\mu$m region is not an artifact from the telluric corrections process.

Extending to longer MIR wavelengths ($\lambda > 5$ $\mu$m) reveals multiple lines of stable Ni in various ionization states (Ni II–IV; J. M. DerKacy et al. 2023; L. A. Kwok et al. 2023; C. Ashall et al. 2024). In particular, the stable [Ni II] 6.636 $\mu$m line has been identified in SN 2021aefx across three epochs of observations (L. A. Kwok et al. 2023; C. Ashall et al. 2024). This feature has also been confidently detected in the underluminous SN 2022xkq at a line velocity of $-460 \pm 110$ km s$^{-1}$, further confirming our findings that blueshifted Ni lines are present in underluminous SNe Ia.

Combining our NIR Ni measurements with recent MIR results may reveal further information about the ionization distribution of the SN ejecta. C. Ashall et al. (2024) measured MIR Ni features in SN 2021aefx and found a larger FWHM in [Ni III] than [Ni II]. JWST spectra of SN 2022xkq show a similar trend of an MIR [Ni III] FWHM of 3790 km s$^{-1}$ as opposed to a FWHM of 2460 km s$^{-1}$ for [Ni II].

When compared to the line widths measured in the NIR, MIR Ni features may reflect an ionization stratification with higher ionization at higher velocities. This corresponds to lower densities and thus lower recombination rates in the outer layers of the ejecta. Therefore, the NIR [Ni II] feature may be a better trace of the stable Ni located in the innermost regions where densities are higher, resulting in more recombination.

### 5.4. Comparisons to Model Predictions

Here, we compare the results of our NIR stable Ni detections to various theoretical predictions concerning progenitor WD mass and different explosion mechanisms. The combination of different central densities, type of burning (i.e., deflagration or detonation), and progenitor metallicity can be used to determine the likelihood of producing stable Ni for different origin scenarios (see Figure 2 of S. Chakraborty et al. 2024 or Figure 1 of P. Hoeflich et al. 2019). In alignment with the initial conditions of the various models discussed below, we do not assume any primordial stable Ni from the progenitor and are only considering stable Ni produced in the explosion. Time-series spectroscopy at late times can also reveal the geometry of the ejecta, and the distribution of stable Ni can be compared to theoretical predictions on the locations of sufficiently high-density regions for different explosion scenarios.

The central density of the progenitor WD depends on the total WD mass, and since the production of stable IGEs requires high-density conditions, detecting stable Ni at late times can possibly distinguish between higher-mass and lower-mass progenitors. Unlike near-$M_{ch}$ models, sub-$M_{ch}$ models often produce no stable Ni (e.g., S. Blondin et al. 2018) and therefore the resulting synthetic spectra exhibit no NIR [Ni II] $\lambda$1.939 $\mu$m feature at both transitional and nebular phases. Therefore, this NIR [Ni II] emission line may be a way to observationally distinguish between sub-$M_{ch}$ and $M_{ch}$ scenarios.

A. Flörs et al. (2020) published optical and NIR spectra of several nebular-phase SNe Ia and reported detections of the NIR [Ni II] $\lambda$1.939 $\mu$m feature in multiple spectra. The detection of stable Ni is not the focus of that study, but the NIR [Ni II] $\lambda$1.939 $\mu$m feature is present in multiple nebular-phase VLT+XShooter spectra of SN 2015F between +181 and +266 days. Our FIRE spectrum of SN 2015F at +68 days does not have a detection of stable nickel, but this is consistent with normal-bright SNe Ia not showing stable Ni features until later times when the ejecta has become optically thin enough to reveal the innermost regions. We reiterate that a "nondetection" via the Gaussian Peak Ratio does not mean the SN has no stable nickel, but the NIR [Ni II] $\lambda$1.939 $\mu$m line is not sufficiently isolated yet to measure the line velocity and width.

SN 2015F is an interesting case because nebular-phase optical spectra show nearly equal strength emission lines of [Fe II] and [Ni II] in the ~7200 Å region after ~+250 days (M. L. Graham et al. 2017). A. Flörs et al. (2020) conclude SN 2015F has a high Ni/Fe mass ratio. However, comparisons to sub-$M_{ch}$ models (I. R. Seitenzahl et al. 2013; K. J. Shen et al. 2018) reveal that unusually high metallicity is required for sub-$M_{ch}$ models to produce enough Ni to match the observations. A. Flörs et al. (2020) favor high-metallicity sub-$M_{ch}$ progenitors for the SNe Ia in their sample; however, the line measurement results of the NIR Ni and Co lines in our sample do not align with this conclusion.

Metallicity is important to the production of stable material because a higher-metallicity progenitor would have an overall lower electron fraction ($Y_e$) due to the higher neutron content, which affects nuclear statistical equilibrium (NSE) throughout the ejecta (P. Höflich et al. 2006). An increase in electron capture at higher densities (such as in the inner regions) shifts NSE from producing radioactive $^{56}$Ni to producing more stable IGEs like $^{58}$Ni (F. Brachwitz et al. 2000; P. Hoeflich et al. 2017). Increasing the metallicity while keeping the density the same would result in even lower $Y_e$, which means the change in NSE from producing radioactive $^{56}$Ni to stable $^{58}$Ni would occur at a lower density, such as in the outer layers.

At nebular phases when $^{56}$Ni has mostly decayed away, Co lines trace the distribution of radioactive $^{56}$Ni in the ejecta, while Ni lines show the distribution of stable IGEs. Fe lines are less conclusive at the phases we are considering (~+50 to +500 days) due to the presence of both radioactive and stable Fe in the ejecta and the longer half-lives of radioactive isotopes like $^{55}$Fe (S. A. Colgate & C. McKee 1969; I. R. Seitenzahl et al. 2009). The Co lines detected in our sample and other published studies (M. L. Graham et al. 2017, 2022; A. Flörs et al. 2018; K. Maguire et al. 2018) are consistently broader than Ni lines, suggesting the shift in NSE from producing radioactive $^{56}$Ni to $^{58}$Ni occurs in the low-velocity, inner regions because the stable Ni is often located at a lower velocity. The combination of low Ni line velocity and Ni features narrower than Co features indicates the production of this stable material occurred in the innermost region, where there are higher rates of electron capture (P. Hoeflich et al. 2021).

It is important to note that the locations of the highest density regions of the SN ejecta may be different for explosion mechanisms such as surface-triggered scenarios like Helium detonation, violent mergers, or head-on collisions. But since the metallicity and therefore $Y_e$ is a global property of the WD progenitor, the effects on NSE would be reflected in the total distribution of stable material in the SN ejecta. In our sample, we consistently detect more narrow stable Ni features when compared to Co lines in the same spectrum, and thus conclude the shift in NSE occurs at a lower velocity, and the SNe Ia in our sample are less likely to be the result of higher-metallicity progenitors.

*5.4.1. Delayed Detonation*

The deflagration phase of the DDT (A. Khokhlov et al. 1992) scenario is of particular interest to this study as it occurs in the innermost regions of the WD and determines the chemical composition of the central region of the SN that is revealed at nebular phases. The DDT scenario begins with carbon ignition via compressional heating in the interior of the progenitor C/O WD. The flame front propagates as a subsonic deflagration that then transitions into a supersonic detonation. DDT models are often associated with near-$M_{ch}$ progenitors because sufficient mass is required to provide enough compressional heating to trigger carbon ignition (P. Hoeflich et al. 2017). Directly detecting signatures of the deflagration phase is difficult at early times because the ejecta is still optically thick and obscuring the central regions, plus the subsequent detonation may destroy possible clues from this first phase of burning (P. Höflich et al. 2006).

The deflagration wave propagates via Rayleigh–Taylor (RT) instabilities along the flame front and starts unbinding the WD (P. Höflich et al. 2006), thus lowering the density. A natural consequence of the turbulence along the deflagration flame front is mixing as the deflagration flame propagates outward. The narrow line widths measured via the [Ni II] 1.939 $\mu$m line indicate stable nickel is confined to a limited region and may be evidence against large-scale mixing. It is possible that the narrow width of the NIR [Ni II] line could constrain the size of the deflagration region, but the subsequent detonation phase may still produce stable Ni at higher velocities if the density is sufficiently high (R. Pakmor et al. 2024).

*5.4.2. Helium Detonation (Double Detonation)*

Sometimes referred to as the double detonation scenario, the He detonation scenario consists of a C/O WD with a surface He shell that detonates and sends a shock wave inward that triggers a secondary detonation in the interior of the WD. The He detonation scenario has spawned many variations, as there may be multiple ways to trigger the initial surface He detonation (e.g., R. Moll & S. E. Woosley 2013; A. Polin et al. 2019).

Since the first carbon ignition occurs near the surface of the progenitor, the He detonation scenario does not require higher progenitor masses to compressionally heat the interior, and thus, He detonation is a popular mechanism for exploding sub-$M_{ch}$ WDs.

In a sub-$M_{ch}$ He detonation scenario, the production of IGEs can occur either in the interior of the progenitor WD or near the surface, depending on where C ignition is triggered by the initial surface He detonation (A. Polin et al. 2021). One consideration for sub-$M_{ch}$ He detonation models is whether the density in the interior of the WD is high enough to synthesize stable IGEs because the central density depends on the mass of the progenitor WD. Multiple sub-$M_{ch}$ He detonation models have successfully produced sufficient radioactive $^{56}$Ni to match observed SN Ia luminosities, but are often unable to reach densities high enough to produce stable Ni (A. Flörs et al. 2018; K. D. Wilk et al. 2018; K. J. Shen et al. 2018).

Two sub-$M_{ch}$ models from K. D. Wilk et al. (2018) are particularly relevant for this study because they consider the production of stable Ni for different burning processes. One model is a pure detonation of a sub-$M_{ch}$ CO WD with the ignition point located in the interior of the WD. The other sub-$M_{ch}$ model is a DDT explosion of an $M_{ch}$ CO WD that has been artificially scaled to produce the same $^{56}$Ni mass and ejecta mass as the first sub-$M_{ch}$ model. The second model does not describe an explosion scenario where a sub-$M_{ch}$ WD produces amounts of $^{56}$Ni comparable to $M_{ch}$ scenarios, but it does offer a chance to compare predictions for deflagration or detonation in the innermost regions of a CO WD.

K. D. Wilk et al. (2018) compute synthetic spectra at +100 days post explosion and find that a pure detonation sub-$M_{ch}$ model produces no [Ni II] $\lambda$1.939 $\mu$m. The $M_{ch}$ and artificially scaled sub-$M_{ch}$ DDT models included in this study both produce a strong NIR [Ni II] feature, indicating that higher central densities and deflagration in the innermost region are required to produce stable IGEs that can be detected at late times.

S. Blondin et al. (2023) compare nebular-phase JWST NIR and MIR spectra to synthetic spectra produced using four different SN Ia models. The $M_{ch}$ models include specific characterizations of certain explosion mechanisms: DDT (I. R. Seitenzahl et al. 2013) and pulsationally assisted gravitationally confined detonation (F. Lach et al. 2022); whereas the lower-mass models include sub-$M_{ch}$ double detonation (S. Gronow et al. 2021) and a merger model of two sub-$M_{ch}$ WDs (R. Pakmor et al. 2021). Interestingly, the [Ni II] 1.939 $\mu$m emission line is only present in the $M_{ch}$ DDT model even though four different explosion mechanisms are considered. Some of these SN Ia models with weak or no NIR [Ni II] exhibit prominent emission lines of higher ionization states like [Ni III] and [Ni IV] that still indicate the presence of stable Ni at late times. Further modeling is required to determine whether the models without the [Ni II] 1.939 $\mu$m feature do not produce any stable Ni in the explosion or if the feature is suppressed by ionization effects (such as those mentioned in S. Blondin et al. 2023) or blending from neighboring lines.

*5.4.3. Mergers and Collisions*

Violent mergers or collisions of two WDs are another way to ignite carbon without relying on compressional heating due to a high progenitor mass. This type of shock-triggered explosion produces regions of rapid compression extreme enough to trigger thermonuclear runaway (D. D. Liu et al. 2016). Observational signatures of violent mergers include significant asymmetries due to the kinematics of the two individual WDs and uneven propagation of the shock front as it starts at the low-density surface of the WDs where the collision occurs (S. Rosswog et al. 2009). Some merger models also consider whether the secondary WD explodes, in which case the ejecta of the primary WD is compressed by the explosion and may push newly synthesized stable Ni into regions with higher density (R. Pakmor et al. 2024). However, simulations have shown that violent mergers may not be viable for all very low-mass WDs (S. Blondin et al. 2023) because SNe Ia require minimum density conditions to synthesize enough radioactive $^{56}$Ni to match observed luminosities (A. J. Ruiter et al. 2013; D. D. Liu et al. 2016).

R. Pakmor et al. (2024) compare several multidimensional violent merger models of two sub-$M_{ch}$ WDs, including prompt mergers and helium-ignited mergers. In the prompt violent merger model of two sub-$M_{ch}$ WDs from (R. Pakmor et al. 2012), only the primary WD produces stable Ni during the initial detonation triggered by the impact. Furthermore, the helium-ignited merger model with both WDs exploding (R. Pakmor et al. 2022) produces very little stable Ni in total, with the majority of that material located at intermediate velocities (∼4000 to ∼10,000 km $s^{-1}$). S. Blondin et al. (2023) use the same violent merger model of two sub-$M_{ch}$ WDs from R. Pakmor et al. (2012) to produce synthetic optical, NIR, and MIR spectra. This merger model had overall lower ionization than other explosion models considered in this work, and the resulting synthetic spectra do not exhibit a [Ni II] $\lambda$1.939 $\mu$m line (see Table 2 of S. Blondin et al. 2023). L. A. Kwok et al. (2024) use the same merger model as S. Blondin et al. (2023) but compute synthetic spectra of the 03fg-like SN Ia 2022pul at a later phase and find [Ni II] $\lambda$1.939 present at +380 days, possibly indicating the NIR Ni feature emerges at later times in this 03fg-like object. Although 2022pul is a different SN Ia subtype than the objects included in our sample, the similarity in behavior of the NIR Ni feature at late times may point to similar physics, such as opacity or ionization structure.

Head-on collisions have also been considered due to their distinct predicted observational signature. If two WDs collide in a head-on collision, they impart a bimodal velocity distribution to the resulting SN ejecta (S. Dong et al. 2015). This will result in double-peaked nebular-phase line profiles as the spectra are a superposition of the spectra of each individual exploding WD.

Based on nebular-phase optical spectra, Vallely et al. (2020) identify SN 2011iv as a bimodal object that may be the result of a head-on collision or merger of two WDs. An underlying bimodal velocity distribution should affect both optical and NIR spectral features, assuming the velocity offset between the two nodes of the bimodal distribution is larger than the spectral resolution of the observations. SN 2011iv has the largest Gaussian Peak Ratio values in this sample and shows a single, narrow peak on the blue side of the 1.9 $\mu$m feature at +54 and +83 days. Based on the NIR [Ni II] line profile, we conclude SN 2011iv is not the result of a head-on collision.

For all SNe Ia in our sample with stable Ni detections, the line profile of the NIR [Ni II] $\lambda$1.939 $\mu$m line exhibits a single narrow peak and is thus unlikely to be the result of head-on collisions or WD–WD mergers (L. A. Kwok et al. 2024; M. R. Siebert et al. 2024). The narrow line widths also rule out the possibility of an underlying bimodal velocity distribution being obscured by the blending of two wider lines.

#### 5.4.4. 1D versus Multidimensional Models

Nebular-phase time-series spectroscopy allows for a more detailed study of the geometric distribution of material along our line of sight, but many theoretical predictions are based on 1D models that may not capture 3D viewing angle effects. R. Pakmor et al. (2024) compare a range of multidimensional SN Ia explosion models and examine the distribution of radioactive and stable Ni shortly after the explosion. A 3D DDT model (I. R. Seitenzahl et al. 2013) predicts the bulk of stable Ni to be at intermediate to high velocities ($\geqslant$6000 km $s^{-1}$), whereas the 1D $M_{ch}$ DDT model (S. Blondin et al. 2013) used for comparison predicts the opposite, with stable Ni confined to the low-velocity, innermost region ($\leqslant$4000 km $s^{-1}$). However, the models used in R. Pakmor et al. (2024) come from multispot ignitions, which produce the intermediate to high velocity of stable Ni. For 3D DDT models of a single spot ignition, it has been shown that stable Ni is centrally located (C. Ashall et al. 2024). Future multidimensional modeling efforts on the effects of buoyancy and RT instabilities along the deflagration front may help resolve current discrepancies between different 3D and 1D model predictions. Based on the observed NIR spectra in our sample, both normal-bright and subluminous 86G-like SNe Ia exhibit centrally located stable Ni at late times.

### 5.5. Subluminous 86G-like SNe Ia

Due to the lower mass of the progenitor WD, some sub-$M_{ch}$ models predict lower yields of radioactive $^{56}$Ni to power the early light curve, resulting in several studies concluding subluminous SNe Ia may be the result of sub-$M_{ch}$ progenitors (e.g., S. Blondin et al. 2018; D. A. Goldstein & D. Kasen 2018). In our sample of SNe Ia, we surprisingly find the strongest stable Ni detections in the two subluminous 86G-like objects: SN 2011iv and SN 2015bp.

Multiple studies have used the relative integrated flux of nebular-phase emission lines to estimate the ratios of stable IGEs produced in SNe Ia (e.g., K. Maguire et al. 2018; A. Flörs et al. 2020; J. Liu et al. 2023). Spectra with the highest Gaussian peak ratio ($r$) values have a 1.9 $\mu$m feature dominated by the [Ni II] 1.939 $\mu$m emission line, so the larger $r$ values of SN 2011iv and SN 2015bp may indicate that more stable Ni was produced in these two SNe Ia compared to other objects in this sample. Since the production of stable Ni requires high density in the central region of the WD, we conclude that the subluminous 86G-like SNe Ia in our sample may be the result of higher-mass ($M_{ch}$) progenitors.

This is consistent with multiple works in which it has been shown that at least some subluminous SNe Ia may come from high-mass DDT explosions (P. Höflich et al. 2002; C. Ashall et al. 2016, 2018; J. M. DerKacy et al. 2024). For example, a growing number of SNe Ia have C detections in their early NIR spectra. One such object is iPTF13ebh, a subluminous 86G-like SN Ia with C absorption features in premaximum optical and NIR spectra (E. Y. Hsiao et al. 2015). iPTF13ebh displays strong C I absorption that weakens as the SN approaches peak brightness. Although iPTF13ebh is an 86G-like, similar early C features have been detected in 91bg-likes such as SN 2022xkq (J. Pearson et al. 2024). The prototypical normal SN Ia SN 2011fe also showed C features in its premaximum spectra, but C features in 2011fe appear to become stronger over time (E. Y. Hsiao et al. 2013).

S. D. Wyatt et al. (2021) searched for early carbon detections in a sample of nine subluminous SNe Ia and found that four of the nine SNe Ia showed strong C detection at early times, and two additional SNe showed tentative C detection. This study includes the two subluminous SNe Ia in our sample: SN 2011iv and SN 2015bp. Similar to iPTF13ebh, S. D. Wyatt et al. (2021) report diminishing optical C features in SN 2015bp at velocities ∼11,000 km $s^{-1}$. SN 2011iv also exhibits C absorption features in premaximum spectra, but the single optical spectrum from S. D. Wyatt et al. (2021) is insufficient to determine whether C features strengthen or diminish over time.

The diminishing NIR C in subluminous 86G-like SNe Ia suggests this may be surviving material from the progenitor WD that is only present in the outermost layers of the SN. This disfavors explosion scenarios such as double detonation because the initial detonation occurs in the He shell at the surface, so little to no C from the progenitor is expected to survive (S. Blondin et al. 2015). Therefore, 86G-likes with early C detections that diminish over time show evidence against a He detonation scenario and are more likely the product of an $M_{ch}$ explosion of a CO WD (E. Y. Hsiao et al. 2015). The strongest stable Ni detections in our sample occurring in subluminous 86G-likes combined with diminishing C absorption at early times provides further observational evidence that 86G-likes may be the result of more massive progenitors.

## 6. Conclusion

Studying the production of stable heavy elements in low redshift SNe Ia will improve our understanding of their progenitor systems and explosion mechanisms. When considering the two leading explosion theories, the location of the initial ignition point is one of the main differences between the DDT scenario and the He detonation scenario. The DDT scenario predicts an initial carbon ignition in the interior of the WD, whereas the He detonation scenario begins with a helium detonation at the surface of the WD. To observationally detect the location of the ignition point, we must examine the outermost and innermost regions of SNe Ia. Therefore, the best opportunities to directly detect the destroyed progenitor WD are at the extremes of an SN Ia's lifetime.

The presence of NIR nickel features at late times may be able to observationally discern between $M_{ch}$ and sub-$M_{ch}$ explosions. Since radioactive $^{56}$Ni synthesized in the explosion decays to $^{56}$Co by late times, Ni emission features are attributed to stable $^{58}$Ni in the inner region of the supernova. Stable $^{58}$Ni is a product of high-density nuclear burning (F. K. Thielemann et al. 1986; K. Iwamoto et al. 1999), so detecting stable Ni may indicate high WD central densities that would require a near-$M_{ch}$ progenitor and thus distinguish between sub-$M_{ch}$ and $M_{ch}$ progenitors.

In this work, we present 79 telluric-corrected NIR spectra of 22 SNe Ia, including 31 previously unpublished nebular-phase spectra. Using a subsample of high-S/N spectra, we use a multi-Gaussian fit of the 1.9 $\mu$m region to detect the presence of the NIR [Ni II] $\lambda$1.939 $\mu$m line in both normal and 86G-like SNe Ia.

The stable Ni detections from this sample provide an observational constraint that can be used to inform SN Ia models:

1. The average measured [Ni II] $\lambda$1.939 $\mu$m line peak velocity is $\sim$1200 km s$^{-1}$.
2. The average measured line width does not exceed $\sim$ 3500 km s$^{-1}$ for all stable Ni detections in this sample.

As the production of stable Ni requires sufficiently high densities, the combination of explosion mechanism and progenitor WD mass must be able to produce high-density conditions in the region between 0 and $\sim$4000 km s$^{-1}$ in velocity space.

In addition to being located at low velocity, most stable $^{58}$Ni emission lines detected in our sample have fairly narrow line widths. These narrow line widths may be observational evidence of stable Ni being restricted to a limited region in velocity space, even at late times (i.e., past +100 days). The narrow line widths disfavor mixing effects that may result in a larger distribution of stable material into the less dense and higher-velocity regions of the SN ejecta. Furthermore, sub-$M_{ch}$ explosion models often require unusually high metallicities to have enough neutron-rich material to form significant amounts of stable $^{58}$Ni (K. J. Shen et al. 2018; A. Flörs et al. 2020). If this neutron-rich material is due to a high progenitor metallicity, then we would expect stable Ni to be produced throughout the SN ejecta and not confined to a narrow region in velocity space as indicated by the measured [Ni II] line widths.

The detection of stable $^{58}$Ni is possible observational evidence for a higher-mass progenitor WD, though the explosion mechanism is also a factor. The production of stable material in the central region is characteristic of a DDT explosion of a massive WD (E. Y. Hsiao et al. 2015). Other observational evidence of subluminous 86G-like SNe Ia as the product of $M_{ch}$ explosions has been reported in previous studies (P. Hoeflich et al. 2021; S. D. Wyatt et al. 2021). When combined with the new evidence presented here using the NIR [Ni II] $\lambda$1.939 $\mu$m line, spectroscopic evidence of subluminous 86G-like SNe Ia strongly suggests they may be the result of $M_{ch}$ DDT explosions.

As the time domain community prepares for the torrent of transient alerts from LSST, we must adapt spectroscopic observing strategies to obtain follow-up observations of new SNe. Compared to other types of transients, SNe Ia are particularly affected by the imminent deluge of observations due to dedicated cosmological surveys obtaining SN Ia light curves for distance measurements. This will lead to a wealth of light-curve information for numerous SNe Ia and will require strategic use of limited spectroscopic resources.

Based on the time evolution of the late-time NIR spectra in this sample, we recommend triggering follow-up spectroscopy to observe the NIR [Ni II] 1.939 $\mu$m line in new SNe Ia based on the subtype or $s_{BV}$ of the object:

1. Subluminous 86G-like or SNe Ia with $s_{BV} \leqslant 0.75$: NIR spectroscopy should be obtained $\sim$+50 to +100 days post-peak brightness
2. Normal-bright SNe Ia with $\sim 0.8 \leqslant s_{BV} \leqslant 1.2$: NIR spectroscopy should be obtained after $\sim$+150 days post-peak brightness

These time frames are based on the time coverage of our sample, and we encourage future observations to test the utility of this suggested follow-up strategy.

It is still unknown if lookback time can contribute to the inherent diversity within the SN Ia population, so developing techniques to directly gauge progenitor properties may help disentangle redshift from the effects of progenitor mass, metallicity, host environment, and explosion mechanism. Future planned programs with JWST and the upcoming Nancy Grace Roman Space Telescope will obtain NIR spectroscopy of high redshift SNe Ia that can be compared to the results of this study.

## Acknowledgments

We thank the Las Campanas technical staff for their continued support over the years. The work of the CSP-II has been generously supported by the National Science Foundation under grants AST-1008343, AST-1613426, AST-1613455, and AST-1613472.

S.K. and M.M. acknowledge support in part from ADAP program grant No. 80NSSC22K0486, from the NSF grant

AST-2206657, and from the HST GO program HST-GO-16656.

C.A. is supported by STScI grants (JWST-GO-02114, JWST-GO-02122, JWST-GO-04522, JWST-GO-03726, JWST-GO-6582, HST-AR-17555, JWST-GO-04217, JWST-GO-6023, JWST-GO-5290, JWST-GO-5057, JWST-GO-6677) and JPL-1717705.

L.G. acknowledges financial support from AGAUR, CSIC, MCIN, and AEI 10.13039/501100011033 under projects PID2023-151307NB-I00, PIE 20215AT016, CEX2020-001058-M, ILINK23001, COOPB2304, and 2021-SGR-01270.

M.D.S. is funded by the Independent Research Fund Denmark (IRFD) via Project 2 grant 10.46540/2032-00022B and by an Aarhus University Nova grant No. AUFF-E-2023-9-28. CSP-II is also supported by a Sapere Aude Level 2 grant funded by the Danish Agency for Science and Technology and Innovation (PI Stritzinger).

J.L. acknowledges support from NSF-2206523 and DOE No. DE-SC0017955.

*Facility:* Magellan–Baade.

*Software:* This work made use of Astropy:[19] a community-developed core Python package and an ecosystem of tools and resources for astronomy (Astropy Collaboration et al. 2013, 2018, 2022).

## Appendix

Here, we show all multi-Gaussian fits for NIR spectra in our sample with $S/N \geqslant 0.5$ in the 1.9 $\mu$m region. The fit method is detailed in Section 3.1. In the figures below (Figures A1—A4), the magenta Gaussian corresponds to [Ni II] 1.939 $\mu$m, the cyan Gaussian is most likely [Co III] 2.0028 $\mu$m, and the overall fit is shown in red. Gaussian Peak Ratios are listed for all spectra and are noted in red for spectra that meet the detection threshold of $r \geqslant 0.5$.

[19] http://www.astropy.org

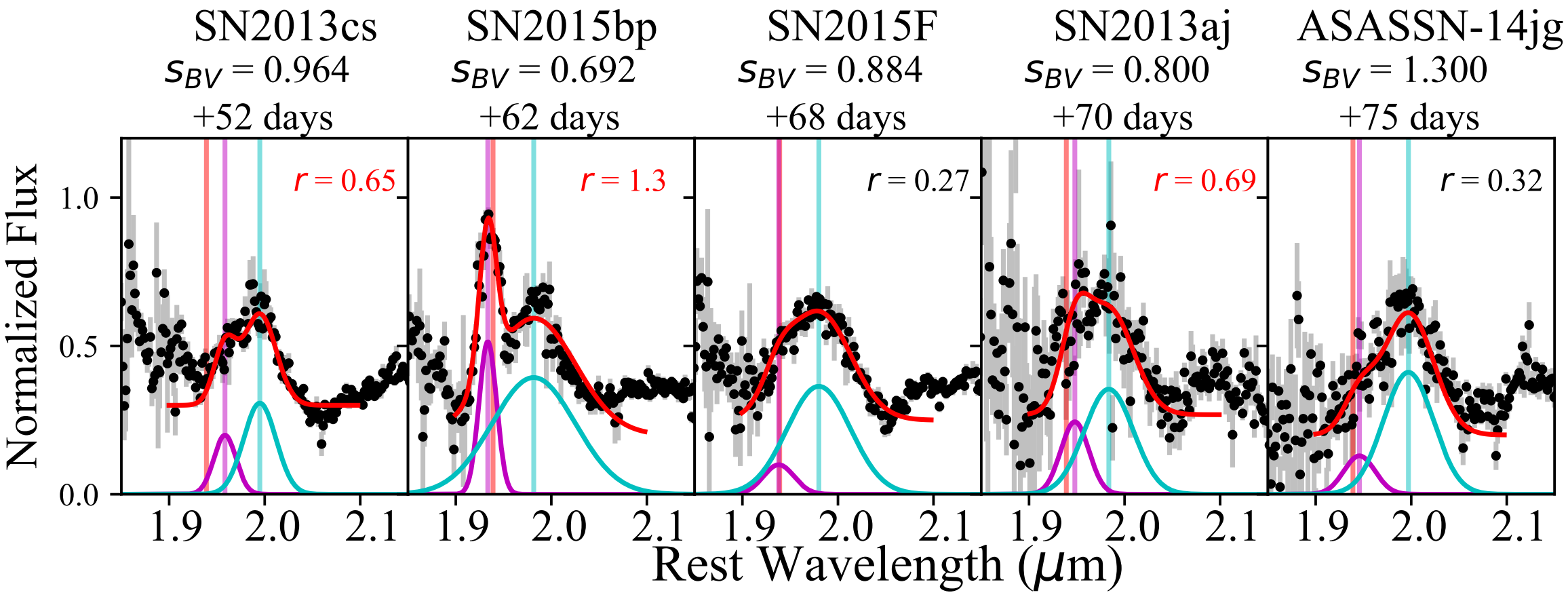

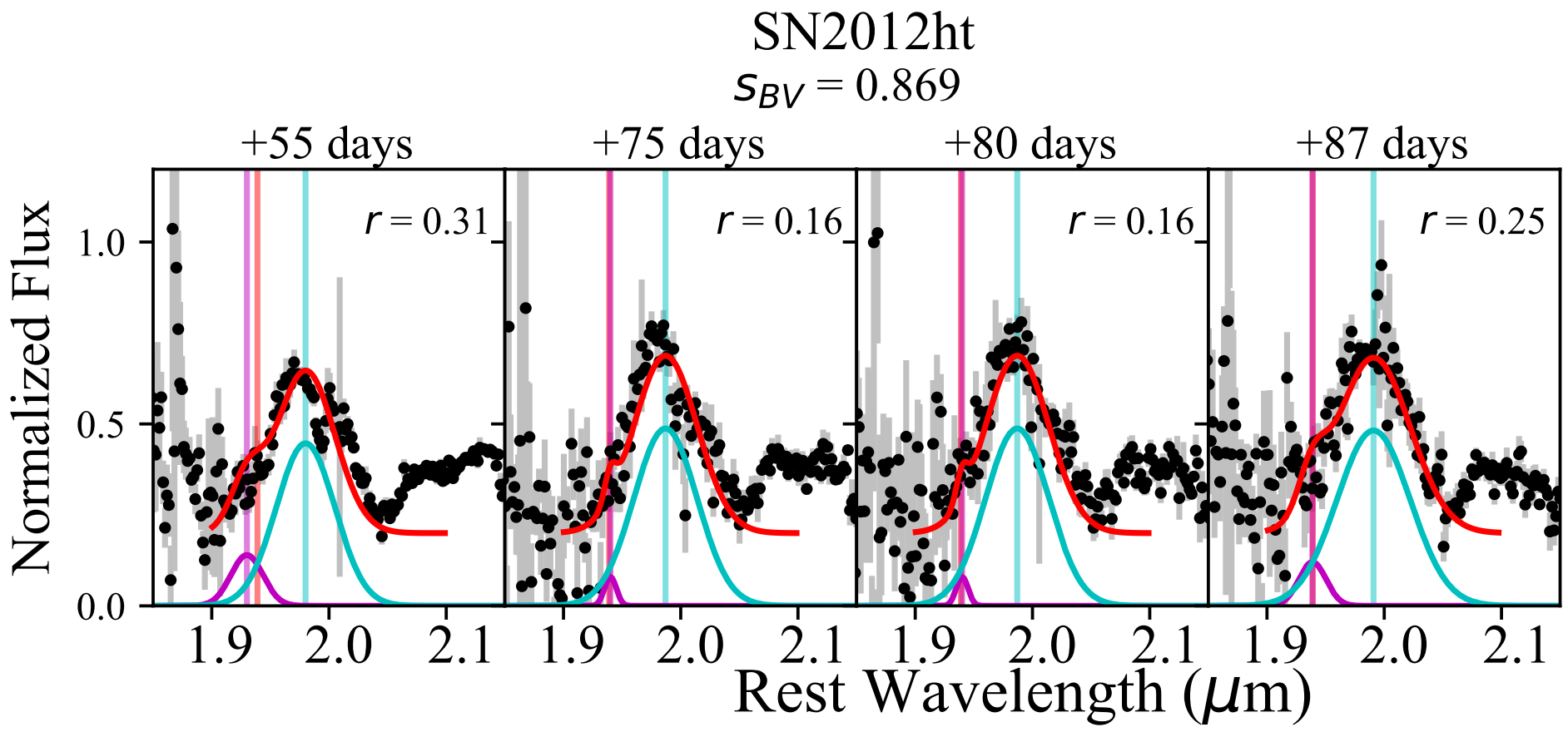

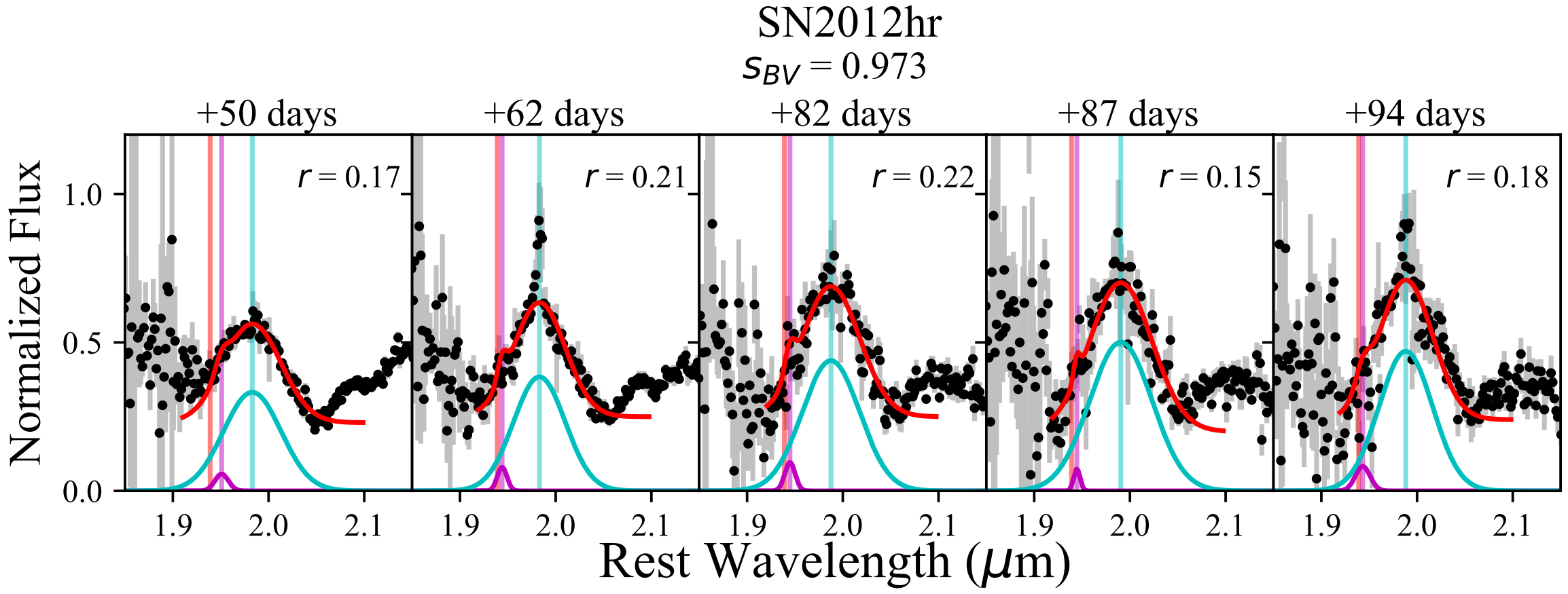

**Figure A1.** Multi-Gaussian fit results for SNe with one spectrum in our sample, as well as SN 2012ht and SN 2012hr.

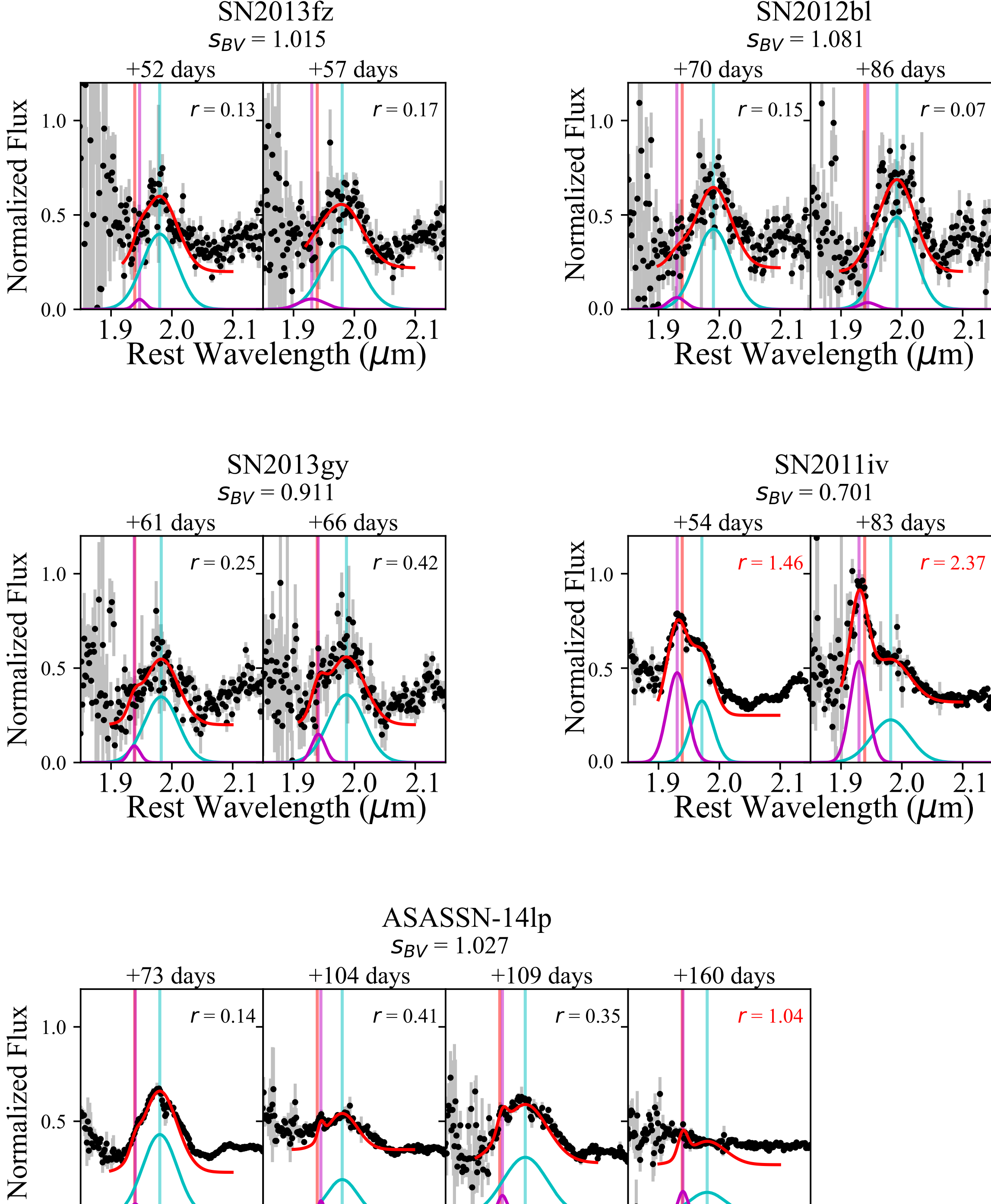

**Figure A2.** Multi-Gaussian fit results for SNe with two spectra in our sample, as well as ASASSN-14lp.

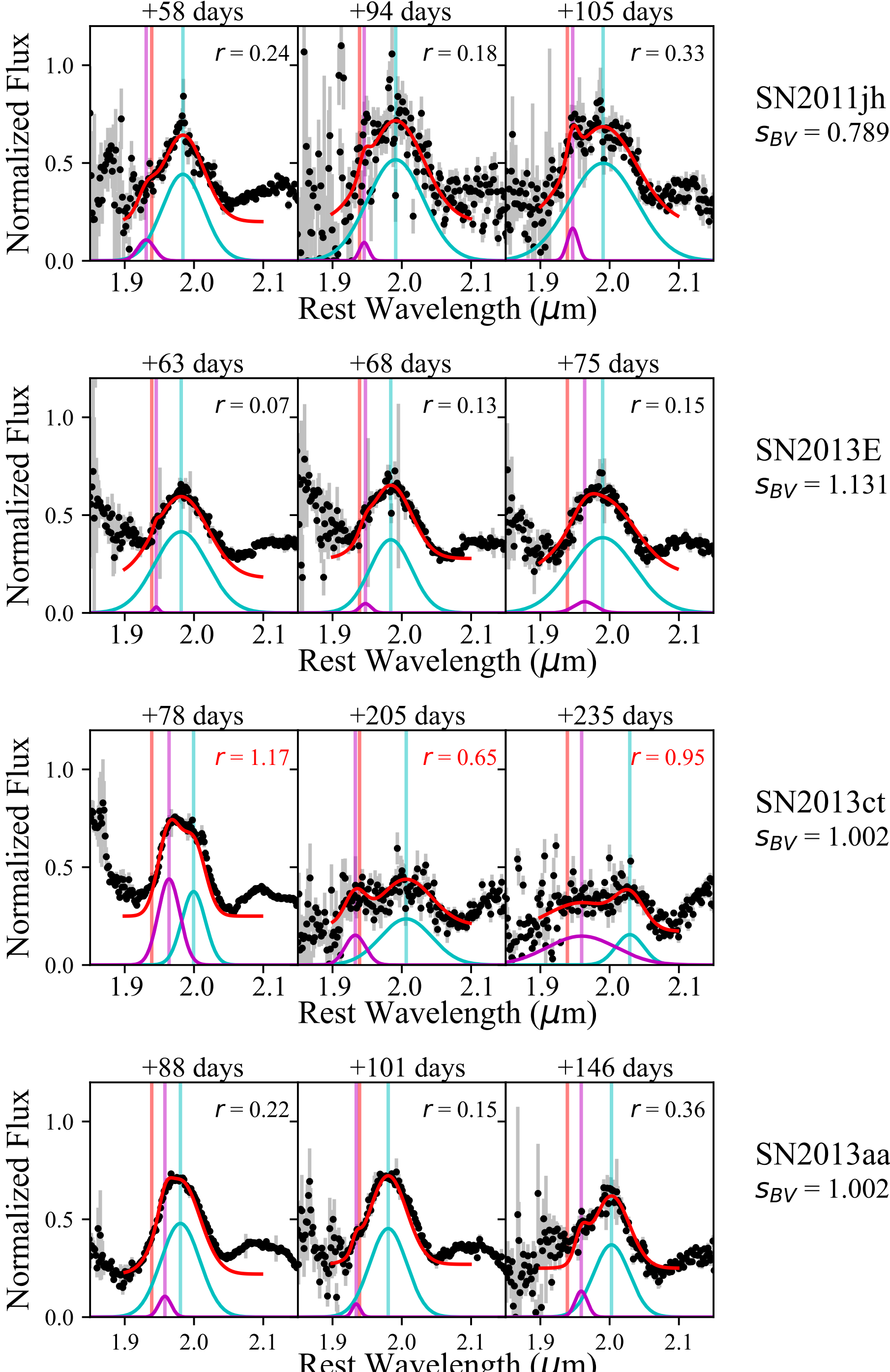

**Figure A3.** Multi-Gaussian fit results of the 1.9 $\mu$m region for SNe in our sample with three NIR spectra.

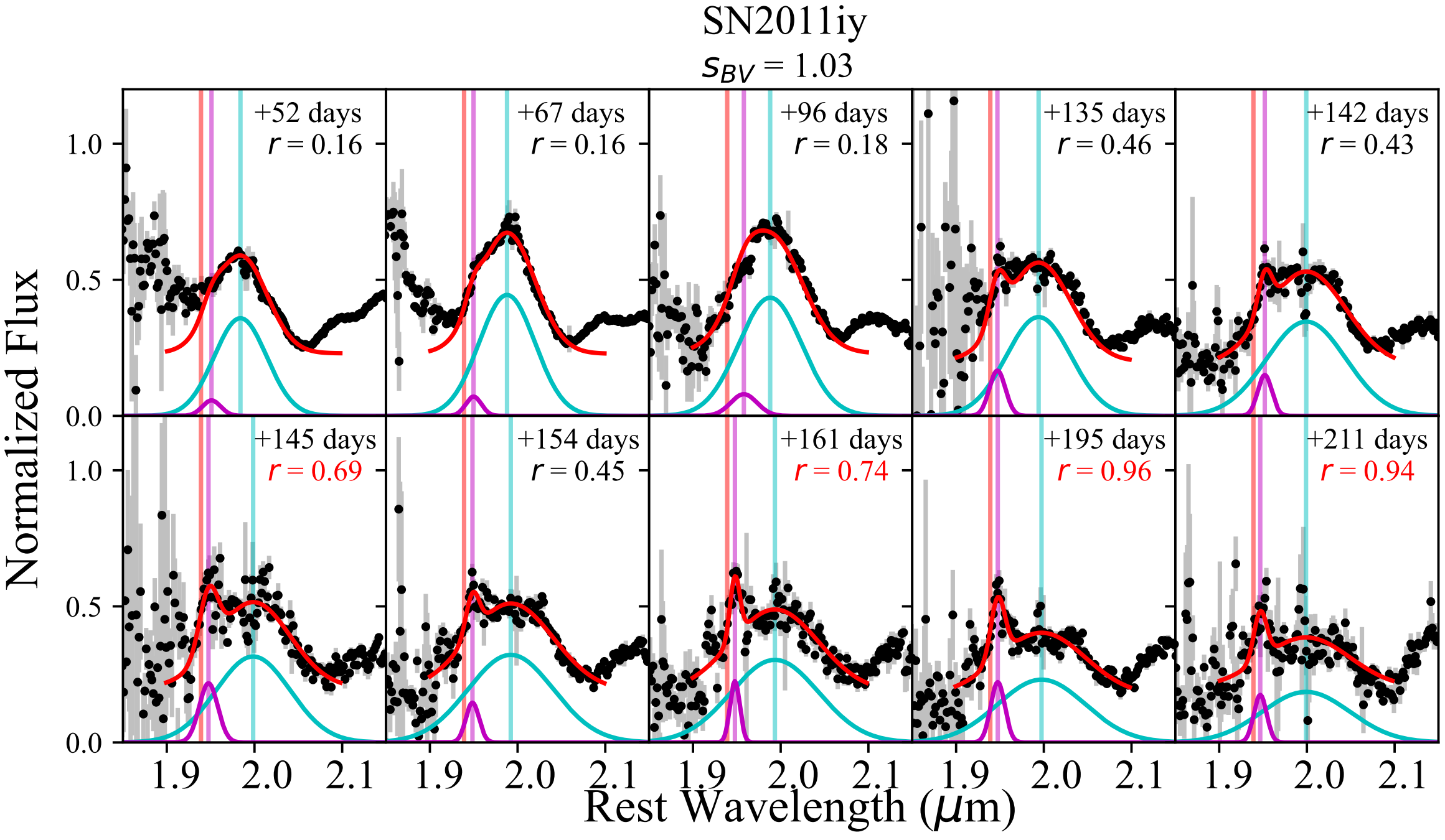

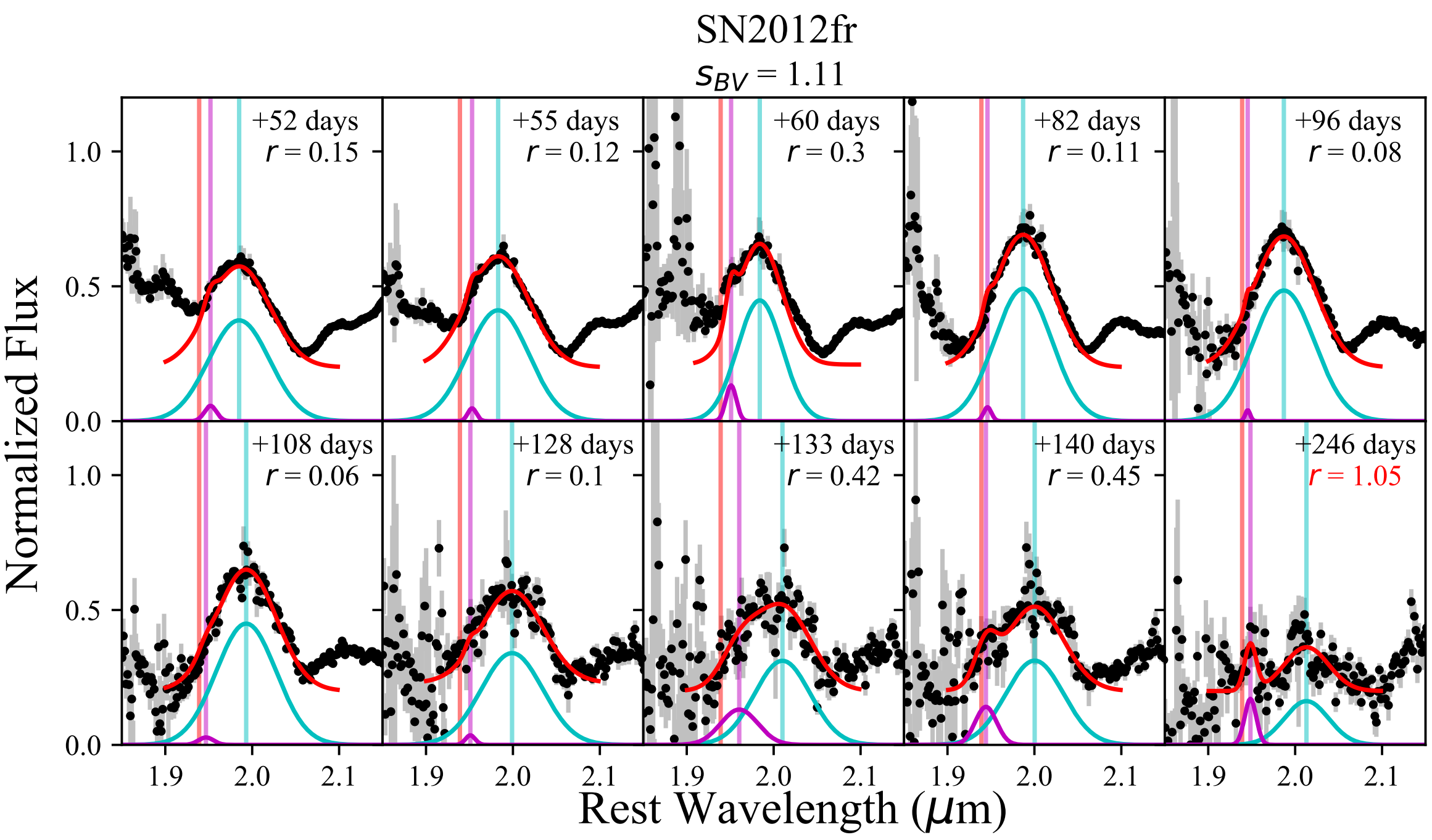

**Figure A4.** Multi-Gaussian fit results for the two SNe in this sample with the most robust NIR spectral time series: SN 2011iy and SN 2012fr.

## ORCID iDs

Sahana Kumar https://orcid.org/0000-0001-8367-7591
Eric Y. Hsiao https://orcid.org/0000-0003-1039-2928
Chris Ashall https://orcid.org/0000-0002-5221-7557
Peter Hoeflich https://orcid.org/0000-0002-4338-6586
Eddie Baron https://orcid.org/0000-0001-5393-1608
Mark M. Phillips https://orcid.org/0000-0003-2734-0796
Maryam Modjaz https://orcid.org/0000-0001-7132-0333
Abigail Polin https://orcid.org/0000-0002-1633-6495
Nidia Morrell https://orcid.org/0000-0003-2535-3091
Christopher R. Burns https://orcid.org/0000-0003-4625-6629
Jing Lu https://orcid.org/0000-0002-3900-1452
Melissa Shahbandeh https://orcid.org/0000-0002-9301-5302
Lindsey A. Kwok https://orcid.org/0000-0003-3108-1328
Lluis Galbany https://orcid.org/0000-0002-1296-6887
M. D. Stritzinger https://orcid.org/0000-0002-5571-1833
Carlos Contreras https://orcid.org/0000-0001-6293-9062
James M. DerKacy https://orcid.org/0000-0002-7566-6080
T. Hoover https://orcid.org/0000-0003-4114-1240
Syed A. Uddin https://orcid.org/0000-0002-9413-4186

Saurabh W. Jha https://orcid.org/0000-0001-8738-6011
Huangfei Xiao https://orcid.org/0009-0006-9436-7197
K. Krisciunas https://orcid.org/0000-0002-6650-694X
N. B. Suntzeff https://orcid.org/0000-0002-8102-181X